\documentclass{article}
\usepackage[table]{xcolor}
\usepackage{booktabs}
\usepackage{multirow}
\usepackage{siunitx}
\usepackage{threeparttable}
\usepackage{PRIMEarxiv}
\usepackage{authblk}
\usepackage[utf8]{inputenc} % allow utf-8 input
\usepackage[T1]{fontenc}    % use 8-bit T1 fonts
\usepackage{hyperref}       % hyperlinks
\usepackage{multirow}       % complex table
\usepackage{url}            % simple URL typesetting
\usepackage{booktabs}       % professional-quality tables
\usepackage{amsfonts}       % blackboard math symbols
\usepackage{amssymb}        % for math symbols like \lesssim
\usepackage{nicefrac}       % compact symbols for 1/2, etc.
\usepackage{microtype}      % microtypography
\usepackage{lipsum}
\usepackage{fancyhdr}       % header
\usepackage{graphicx}       % graphics
\usepackage{comment}        % comment out
\usepackage{mathtools}
\usepackage{float}
\usepackage{adjustbox}
\usepackage{bm}
\title{FlowMesh: Deterministic Generative Transport of Physical Fields on Unstructured Meshes}

\author[1]{Siddharth Verma}
\author[2,3]{Jaewan Park}
\author[4]{Jay Phil Yoo}
\author[4]{Kazuma Kobayashi}
\author[5,6]{Souvik Chakraborty}
\author[2,4,*]{Syed Bahauddin Alam}

\affil[1]{Department of Mechanical Engineering, Indian Institute of Technology Bombay, Mumbai, India
}

\affil[2]{National Center for Supercomputing Applications, Urbana, IL, USA
}

\affil[3]{Grainger College of Engineering, Mechanical Science \& Engineering Department, University of Illinois Urbana-Champaign, Urbana, IL, USA
}

\affil[4]{Grainger College of Engineering, Nuclear, Plasma \& Radiological Engineering Department, University of Illinois Urbana-Champaign, Urbana, IL, USA
}

\affil[5]{Department of Applied Mechanics, Indian Institute of Technology Delhi, New Delhi, India
}

\affil[6]{Yardi School of Artificial Intelligence, Indian Institute of Technology Delhi}

\affil[*]{Corresponding author}

\begin{document}
\maketitle

% \begin{abstract}

% Scientific generative models must capture distributions of physical fields on unstructured meshes, yet most architectures assume regular grids or rely on costly stochastic sampling. We introduce FlowMesh, a mesh-native generative framework that couples a permutation-invariant PointNet autoencoder with conditional flow matching and an $\ell_1$ latent-velocity penalty. Deterministic ODE transport generates full-field samples without iterative denoising. We evaluate FlowMesh on lid-driven-cavity, pressurized-water-reactor subchannel and heat-exchanger flows using six fixed, matched seeds. FlowMesh increases macro-averaged probability-density-function coverage by up to 3.77 percentage points over otherwise identical unregularized flow matching and by 45.37 points over a latent diffusion model. Regularization is neutral on the already accurate cavity case but improves distributional fidelity for every reactor and heat-exchanger field, with the largest gains in temperature. These results demonstrate that velocity-regularized deterministic latent transport provides an accurate and computationally efficient route to distributional modeling of physical fields on fixed unstructured meshes.

% \end{abstract}
\begin{abstract}
Generating distributions of high-dimensional physical fields is central to uncertainty quantification in energy systems, but ensemble computational-fluid-dynamics simulations remain prohibitively expensive and are commonly defined on unstructured meshes. Here we introduce \textit{FlowMesh}, a generative framework that combines permutation-invariant mesh encoding with deterministic latent flow matching and an $\ell_1$ penalty on the learned transport velocity. FlowMesh generates full-field realizations on fixed unstructured meshes without interpolation to an auxiliary Cartesian grid or an iterative stochastic denoising chain. Across lid-driven-cavity, pressurized-water-reactor subchannel, and heat-exchanger datasets evaluated over six matched training seeds, FlowMesh achieves macro-averaged probability-density-function coverage of $91.7\%$--$97.0\%$, exceeding a latent DDPM baseline built on the same PointNet representation by $11.6$--$45.4$ percentage points. The effect of velocity regularization is regime dependent: it is negligible in the cavity benchmark, but increases macro-averaged coverage by $3.8$ percentage points for the PWR subchannel and $1.1$ points for the heat exchanger. Autoencoder-only reconstruction exceeds $99.4\%$ coverage in the PWR case, locating most remaining error in latent transport. Each FlowMesh realization requires $1.23 \pm 0.15\,\mathrm{s}$, corresponding to a $585$--$1{,}902$-fold wall-clock acceleration over the reference CFD calculations. These results establish deterministic latent transport as an accurate and computationally efficient alternative to standard latent diffusion.

\end{abstract}

% keywords can be removed
\keywords{Scientific Machine Learning (SciML) \and Generative AI \and Flow Matching \and Unstructured Meshes \and Computational Fluid Dynamics (CFD) \and AI for Energy Systems}

\section*{Introduction}

Many decisions in physical and energy systems depend not on a single best-estimate field but on a distribution of plausible states. Nuclear thermal-hydraulic analysis, heat-exchanger design, virtual sensing, and uncertainty-aware digital twins require ensembles of velocity, temperature, pressure, and turbulence fields over uncertain operating conditions. High-fidelity finite-element and computational-fluid-dynamics solvers \cite{zienkiewicz2005finite,anderson1995computational} provide these fields by resolving the governing conservation equations \cite{quarteroni1994numerical}, but generating sufficiently large ensembles remains computationally expensive \cite{benner2015survey}. This creates a fundamental scale mismatch between the cost of high-fidelity simulation and the number of realizations required for uncertainty propagation, probabilistic analysis, and near-real-time state estimation \cite{kobayashi2024deep,yao2024uncertainty}.

A second difficulty is geometric. Engineering simulations are commonly defined on unstructured meshes whose nonuniform resolution conforms to curved boundaries, narrow flow passages, wall-adjacent gradients, and locally refined regions \cite{frey2007mesh,brostein2017geometric}. In reactor subchannels and heat-transfer components, these geometric features are part of the physical problem rather than a numerical inconvenience \cite{thompson1984grid,mavriplis1999directional}. Mapping such fields to auxiliary Cartesian grids can introduce interpolation error and alter the representation of precisely the regions in which gradients are largest \cite{thompson1998handbook,sanchez2020learning}. A useful scientific generative model should therefore operate on the computational representation on which the high-fidelity data are produced.

Diffusion models provide a powerful route to high-dimensional generative modeling, but their standard formulation introduces an additional computational question. Samples are generated through an iterative reverse process that reconstructs data from progressively corrupted states \cite{ho2020denoising,rombach2022high}. This strategy is highly successful in image generation, yet scientific field generation raises a different requirement: whether comparable or better distributional fidelity can be obtained through a continuous deterministic transport map while avoiding a long denoising chain. The distinction is especially relevant on learned latent representations of irregular meshes, where the geometry encountered by the generative model is induced by the encoder rather than supplied explicitly to the sampling process.

Iterative denoising can accumulate approximation and discretization error, and accelerated samplers may move intermediate states into regions in which the learned denoiser is less accurate \cite{tang2026contractive,yuzhe2025manifold}. These observations do not imply that stochastic generative models are intrinsically incompatible with physical systems. Rather, they motivate a testable question: when the target data are high-dimensional physical fields represented on an irregular mesh, does deterministic transport through the learned latent space provide a more effective generative mechanism than iterative stochastic denoising? We address this question with an architecture-controlled comparison in which FlowMesh and a latent DDPM share the same PointNet autoencoder and latent training data. This design separates the representation problem from the generative-transport problem and allows the effect of the sampling mechanism to be assessed directly on the present benchmarks.
\begin{figure}[htbp]
    \centering
    \includegraphics[width=1\textwidth]{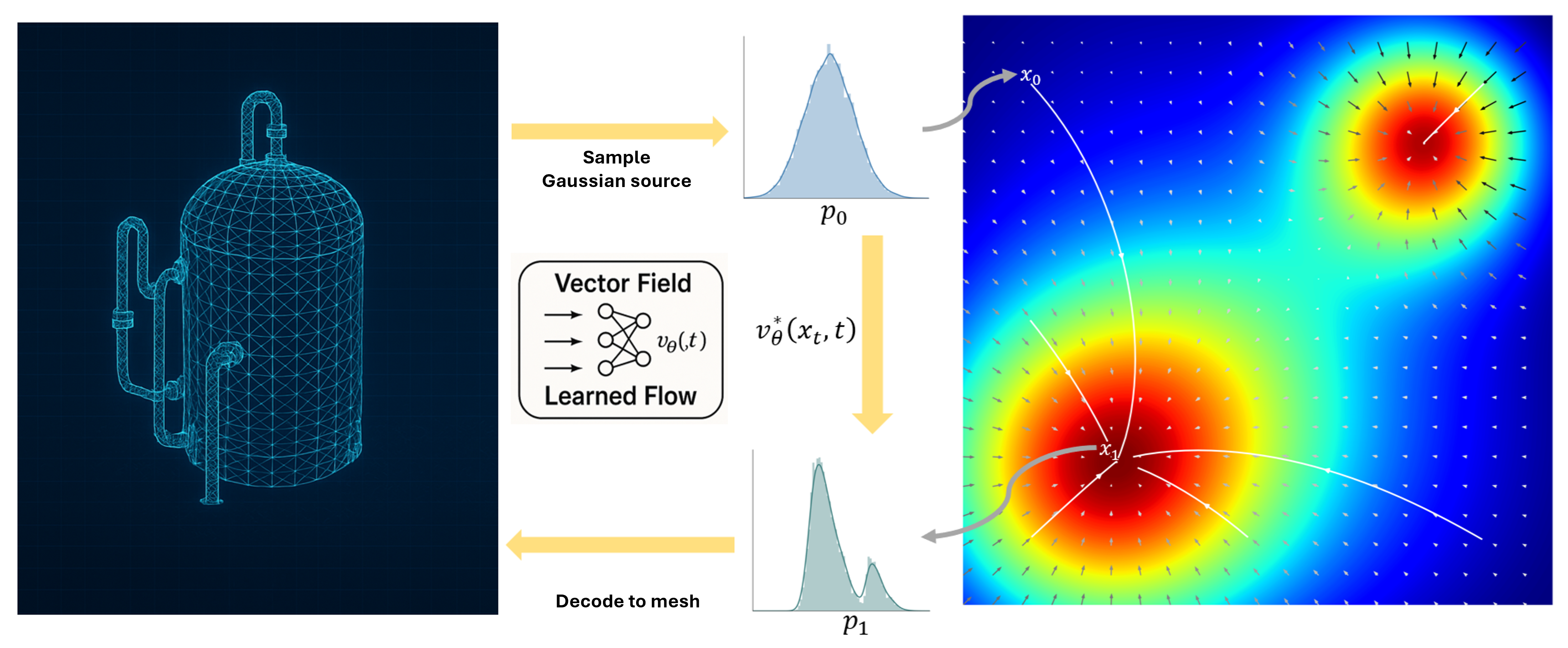}
\caption{\textbf{FlowMesh formulates mesh-based field generation as deterministic latent transport.}
A permutation-invariant PointNet autoencoder maps physical fields and node coordinates from a fixed unstructured mesh into a compact latent representation. A learned time-dependent velocity field $\mathbf{u}_\theta(\mathbf{z}_t,t)$ then transports samples from a Gaussian source distribution $p_0$ to the latent data distribution $p_1$ through numerical integration of an ordinary differential equation. The decoder maps the terminal latent state back to the original mesh without interpolation to an auxiliary Cartesian grid. The schematic contrasts this continuous transport formulation with iterative denoising; it is conceptual and does not imply that individual learned trajectories coincide with a physical trajectory or an optimal-transport geodesic.}
    \label{fig:flow_matching_nuclear}
\end{figure}

Here, we introduce \textbf{FlowMesh}, a generative framework for learning distributions of physical fields on fixed unstructured geometries through deterministic latent transport. Three design choices define the approach. First, a permutation-invariant PointNet autoencoder \cite{qi2017pointnet} jointly embeds node coordinates and node-associated physical quantities into a fixed-dimensional latent representation without interpolation onto an auxiliary grid. Second, flow matching \cite{lipman2022flow} learns a continuous time-dependent velocity field $\mathbf{u}_\theta(\mathbf{z}_t,t)$ whose ordinary differential equation transports samples from a Gaussian reference distribution to the latent distribution of the training fields. Third, an $\ell_1$ penalty on the latent velocity magnitude provides a tunable bias against unnecessarily large transport velocities. This penalty is not a Dirichlet-energy term and does not impose a formal Lipschitz constraint. Instead, its effect is evaluated empirically through a five-order-of-magnitude sweep in $\lambda$, together with measurements of distributional accuracy and the sampled Jacobian norm of the learned velocity field. The resulting experiments allow us to distinguish where velocity regularization materially improves latent transport from regimes in which it is unnecessary or detrimental.

FlowMesh addresses a different objective from the deterministic neural operators commonly used for forward surrogate modeling, including DeepONet, MIONet, S-DeepONet, the Fourier neural operator, and transformer-based operators \cite{lu2021learning,jin2022mionet,he2024sequential,li2020fourier,hao2023gnot}. In their standard supervised form, these models learn a mapping from prescribed inputs to a point prediction of the corresponding field. FlowMesh instead targets the distribution of full fields represented in a simulation ensemble. In the formulation studied here, this distribution is learned over the sampled operating envelope rather than conditioned on a user-specified operating point. The resulting samples are therefore suited to ensemble characterization and uncertainty-propagation tasks rather than deterministic forward prediction. Probabilistic neural operators and conditional generative operators represent an important adjacent class of methods; direct comparison with those formulations is outside the present scope. We therefore treat FlowMesh as complementary to deterministic operator learners rather than as a replacement for them.

We evaluate FlowMesh on three benchmarks that progressively increase the geometric and thermal complexity of the generated fields: a lid-driven cavity, a pressurized-water-reactor subchannel, and a heat exchanger. Six matched training seeds establish that velocity regularization is effectively neutral on the already accurate cavity benchmark but improves the mean distributional fidelity of the PWR and heat-exchanger ensembles, with the largest gain occurring for temperature. An autoencoder-only diagnostic reaches more than $99.4\%$ PDF coverage in the PWR case, showing that the dominant remaining error arises in latent distribution transport rather than mesh reconstruction. Against a latent DDPM using the same PointNet representation, FlowMesh improves macro-averaged PDF coverage by $11.6$--$45.4$ percentage points and reduces sampling cost. Relative to the reference CFD calculations, each generated realization is $585$--$1{,}902$ times faster. At the same time, a no-slip diagnostic reveals a substantial boundary-condition residual, establishing an important boundary of the present result: high distributional fidelity does not by itself guarantee pointwise satisfaction of the governing constraints. Together, the experiments identify deterministic latent transport as a viable generative representation for physical-field ensembles on fixed unstructured meshes while showing that explicit physical constraints remain necessary for safety-critical deployment.

\section*{Results}
\paragraph{Choice of benchmarks} The three benchmarks are selected to test distinct aspects of the generative problem rather than to reproduce industrial scale. The lid-driven cavity provides a comparatively simple recirculating flow in which unregularized flow matching is already accurate and therefore acts as a useful negative control for the regularizer. The PWR subchannel introduces nuclear-relevant rod geometry, heated boundaries, and coupled velocity, turbulence, and temperature fields. The heat exchanger tests whether the observed behavior persists on a second energy-system geometry with a larger set of thermofluid variables. Each dataset retains the native fixed unstructured node set used by the CFD calculation. The objective is therefore narrower than universal mesh generalization: we test whether deterministic latent transport can reproduce distributions of high-dimensional fields on fixed irregular geometries, and whether velocity regularization provides a reproducible benefit when unregularized transport is less accurate.
% (1) LDC
\subsection*{Lid-driven cavity flow}

We evaluate FlowMesh on the canonical lid-driven cavity (LDC) benchmark to assess its generative performance on an advection-dominated flow. Consider a two-dimensional lid-driven cavity defined on $\Omega=[0,1]^2$. A prescribed time-dependent velocity drives the upper boundary in the horizontal direction, while zero-velocity conditions are imposed on the remaining walls. The flow fields are obtained from the incompressible Reynolds-averaged Navier-Stokes equations with a $k$--$\varepsilon$ turbulence closure. Denoting the velocity by $\boldsymbol{v}=(v_x,v_y)$ and the pressure by $p$, the equations of continuity and momentum are:

\begin{equation}
\nabla \cdot \boldsymbol{v} = 0,
\end{equation}
\begin{equation}
\frac{\partial \boldsymbol{v}}{\partial t} + (\boldsymbol{v}\cdot\nabla)\boldsymbol{v} 
= -\nabla p + \nabla \cdot \Big[ (\nu + \nu_t)\big( \nabla \boldsymbol{v} + \nabla \boldsymbol{v}^\top \big) \Big],
\end{equation}
where $\nu$ is the kinematic viscosity and $\nu_t$ the eddy viscosity.  

The turbulent kinetic energy (TKE) field $k(x,y,t)$ evolves as
\begin{equation}
\frac{\partial k}{\partial t} + \boldsymbol{v}\cdot\nabla k 
= P_k - \varepsilon + \nabla \cdot \Big[ \Big( \nu + \tfrac{\nu_t}{\sigma_k} \Big) \nabla k \Big],
\end{equation}
where $P_k$ is the production term, $\varepsilon$ the dissipation rate, and $\sigma_k$ a model constant.  

\begin{figure}[htbp]
    \centering
    \includegraphics[width=0.99\textwidth]{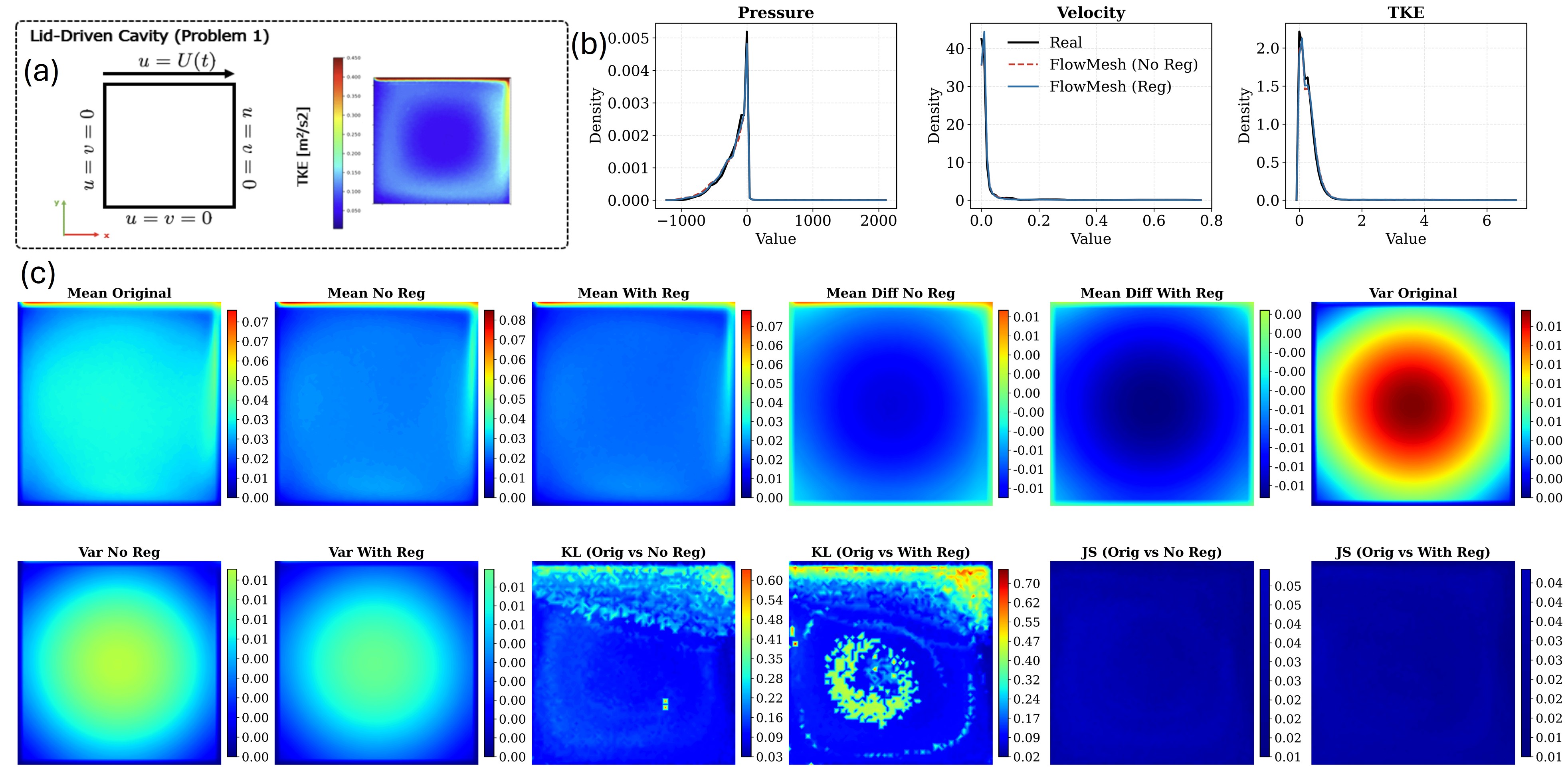}
    \caption{
    Lid-driven cavity benchmark results on an unstructured mesh.
    \textbf{(a)} Two-dimensional lid-driven cavity configuration with prescribed boundary conditions and the turbulent kinetic energy (TKE) field at the terminal simulation time.
    \textbf{(b)} Probability density function (PDF) comparisons between the reference computational fluid dynamics (CFD) data and FlowMesh-generated samples for pressure, velocity magnitude, and TKE.
    \textbf{(c)} Node-wise diagnostics for velocity magnitude, including mean and variance fields from the reference data and generated samples with and without regularization, mean-error maps relative to the reference, and Kullback-Leibler (KL) and Jensen-Shannon (JS) divergence maps between the reference and generated node-wise distributions.
    }
    \label{fig:ldc_case}
\end{figure}

\begin{table}[h]
  \centering
  \caption{LDC $\ell_1$ velocity-regularization sweep across \textbf{6 independent training seeds} (PDF coverage, mean $\pm$ standard deviation, evaluated against autoencoder-reconstructed reference fields). Coverage varies by less than approximately one percentage point for $0\leq\lambda\leq10^{-1}$, whereas $\lambda=1$ produces a large loss of fidelity in all three fields. The result establishes a broad low-to-moderate regularization regime in which the cavity solution is insensitive to $\lambda$, followed by clear over-regularization at the largest weight.}
  \label{tab:ldc_sweep}
  \small
  \begin{tabular}{l c c c}
    \toprule
    $\lambda$ & \textbf{Pressure} & \textbf{Velocity} & \textbf{TKE} \\
    \midrule
    $0$ & $96.0 \pm 1.6$ & $95.7 \pm 1.0$ & $97.7 \pm 1.3$ \\
    $10^{-4}$ & $96.2 \pm 2.0$ & $95.4 \pm 0.9$ & $97.7 \pm 1.4$ \\
    $10^{-3}$ & $96.2 \pm 1.2$ & $95.7 \pm 0.8$ & $97.8 \pm 1.2$ \\
    $10^{-2}$ & $96.4 \pm 1.4$ & $95.7 \pm 0.8$ & $97.6 \pm 1.2$ \\
    $10^{-1}$ & $96.8 \pm 0.9$ & $96.0 \pm 0.7$ & $98.1 \pm 0.8$ \\
    $1.0$ & $50.8 \pm 1.5$ & $14.9 \pm 0.1$ & $60.1 \pm 3.4$ \\
    \bottomrule
  \end{tabular}
\end{table}

\begin{table*}[t]
  \centering
  \begin{threeparttable}
    \caption{LDC distributional metrics over 6 training
    seeds (mean $\pm$ standard deviation). All configurations share the same
    architecture, data, and 100-epoch training schedule. Coverage is evaluated
    against autoencoder-reconstructed reference fields. Regularization changes
    coverage by less than one percentage point, within the seed-to-seed
    variability. Lower KL and JS and higher PDF coverage are better.}
    \label{tab:ldc_seed_metrics}
    \vspace{4pt}

    \small
    \setlength{\tabcolsep}{3.5pt}
    \renewcommand{\arraystretch}{1.08}

    \begin{tabular}{@{}llccc@{}}
      \toprule
      \textbf{Model}
      & \textbf{Field}
      & \textbf{KL}~$\downarrow$
      & \textbf{JS}~$\downarrow$
      & \textbf{PDF Coverage (\%)}~$\uparrow$ \\
      \midrule

      \multirow{3}{*}{%
        \shortstack[l]{FlowMesh $\ell_1$\\No Reg.}}
      & Pressure
      & $0.0071 \pm 0.0058$
      & $0.0027 \pm 0.0024$
      & $96.0 \pm 1.6$ \\

      & Velocity
      & $0.0173 \pm 0.0063$
      & $0.0054 \pm 0.0017$
      & $95.7 \pm 1.0$ \\

      & TKE
      & $0.0027 \pm 0.0019$
      & $0.0010 \pm 0.0007$
      & $97.7 \pm 1.3$ \\
      \midrule

      \multirow{3}{*}{%
        \shortstack[l]{FlowMesh $\ell_1$, Reg.\\($\lambda=10^{-3}$)}}
      & Pressure
      & $0.0060 \pm 0.0041$
      & $0.0023 \pm 0.0016$
      & $96.2 \pm 1.2$ \\

      & Velocity
      & $0.0180 \pm 0.0071$
      & $0.0057 \pm 0.0020$
      & $95.7 \pm 0.8$ \\

      & TKE
      & $0.0023 \pm 0.0018$
      & $0.0008 \pm 0.0006$
      & $97.8 \pm 1.2$ \\
      \midrule

      \multirow{3}{*}{%
        \shortstack[l]{FlowMesh $\ell_1$, Reg.\\($\lambda=10^{-1}$)}}
      & Pressure
      & $0.0057 \pm 0.0030$
      & $0.0018 \pm 0.0009$
      & \cellcolor{gray!15}$\bm{96.8 \pm 0.9}$ \\

      & Velocity
      & $0.0178 \pm 0.0074$
      & $0.0055 \pm 0.0019$
      & \cellcolor{gray!15}$\bm{96.0 \pm 0.7}$ \\

      & TKE
      & $0.0017 \pm 0.0009$
      & $0.0006 \pm 0.0003$
      & \cellcolor{gray!15}$\bm{98.1 \pm 0.8}$ \\
      \bottomrule
    \end{tabular}
  \end{threeparttable}
\end{table*}

% ($\lambda{=}10^{-1}$)
\begin{table*}[h]
  \centering
  \begin{threeparttable}
    \caption{LDC (lid-driven cavity): comparison of statistical values of data across various fields. FlowMesh rows are the mean over \textbf{6 independent training seeds} (same architecture, data and 100-epoch schedule; only the training seed differs); the Original row is as reported for the original dataset. Negative velocity-magnitude \& TKE values are physically inadmissible; the table reports raw model outputs prior to any non-negativity post-processing.}
    \label{tab:LDC_model_stats_seed}
    \vspace{4pt}
    \begin{tabular}{
      l
      l
      S[table-format=-4.4]
      S[table-format=-4.4]
      S[table-format=-4.4]
    }
      \toprule
      \textbf{Data Origin} & \textbf{Field} & {\textbf{Mean}} & {\textbf{Min Value}} & {\textbf{Max Value}} \\
      \midrule
      \multirow{3}{*}{Original}
        & Pressure                  & -234.7000 & -1570.1500 & 2515.7000 \\
        & Velocity                  & 0.0320 & 0.0000 & 0.7470 \\
        & TKE & 0.3130 & 0.0000 & 7.4450 \\
      \midrule
      \multirow{3}{*}{FlowMesh No Regularizer}
        & Pressure                  & -226.9062 & -1239.5104 & 2140.1989 \\
        & Velocity                  & 0.0206 & -0.0051 & 0.7677 \\
        & TKE & 0.3086 & -0.1371 & 6.9881 \\
      \midrule
      \multirow{3}{*}{FlowMesh With Regularizer ($\lambda{=}10^{-1}$)}
        & Pressure                  & -207.8638 & -1179.5140 & 2037.2308 \\
        & Velocity                  & 0.0203 & -0.0054 & 0.7500 \\
        & TKE & 0.2966 & -0.1503 & 6.7631 \\
      \bottomrule
    \end{tabular}
  \end{threeparttable}
\end{table*}

Boundary conditions are given by
\begin{equation}
v_x(x,1,t) = V(t), \quad v_y(x,1,t)=0, \qquad x \in [0,1],
\end{equation}
\begin{equation}
\boldsymbol{v} = (0,0) \quad \text{on all remaining walls}.
\end{equation}
A uniform pressure field and a zero velocity was set as a initial condition for this system:
\begin{equation}
\boldsymbol{v}(x,y,0) = (0,0), \quad p(x,y,0) = p_0.
\end{equation}

We generate the lid velocity $V(t)$ randomly and discretize it into $90$ uniform time steps, serving as the input forcing profile. Simulations are performed in \textsc{ANSYS Fluent} using an unstructured mesh with $4225$ spatial nodes. The numerical solver outputs the velocity magnitude $\|\boldsymbol{v}\| = \sqrt{v_x^2+v_y^2}$, the pressure field $p(x,y,t)$, and the TKE field $k(x,y,t)$ at every timestep. At the terminal simulation time, these fields constitute the training dataset. We trained FlowMesh on 20{,}000 samples drawn from the full simulation dataset. Generated and reference fields are compared in Fig.~\ref{fig:ldc_case}. 

Without regularization, FlowMesh attains PDF coverages of $97.7\pm1.3\%$ for TKE, $95.7\pm1.0\%$ for velocity magnitude, and $96.0\pm1.6\%$ for pressure, with corresponding KL divergences of $0.0027\pm0.0019$, $0.0173\pm0.0063$, and $0.0071\pm0.0058$, respectively (Table~\ref{tab:ldc_seed_metrics}). Node-wise statistics (Fig.~\ref{fig:ldc_case}c) show that the mean and variance are reproduced across most of the domain, with the largest discrepancies concentrated near the moving lid and the bottom corners, where the flow gradients are strongest.

At $\lambda=10^{-1}$, regularization increases the mean PDF coverage by only 0.8, 0.3, and 0.4 percentage points for pressure, velocity, and TKE, respectively. These changes are smaller than the corresponding seed-to-seed standard deviations, and the reported mean-$\pm$-standard-deviation intervals overlap for all three fields. The divergence metrics are also field dependent: KL and JS decrease modestly for pressure and TKE but change slightly unfavorably for velocity. The results therefore do not establish a clear LDC improvement from regularization. The summary statistics in Table~\ref{tab:LDC_model_stats_seed} provide a complementary assessment. The unregularized model is closer to the original data in eight of the nine reported mean, minimum, and maximum statistics, whereas the regularized model is closer only for the maximum velocity. This does not contradict the marginal increase in PDF coverage, because coverage measures overlap of the full empirical distribution, while the reported means and extrema probe only selected statistics and are particularly sensitive to tail samples.

Taken together, the LDC benchmark provides an important negative control. Unregularized flow matching already attains high distributional fidelity, and the changes produced by $\lambda\leq10^{-1}$ are smaller than the seed-to-seed variability for the principal coverage metrics. The data therefore do not support a claim that regularization universally improves FlowMesh. Instead, they show that a broad interval of low-to-moderate regularization leaves an already accurate transport model largely unchanged. At $\lambda=1$, coverage collapses to $50.8\%$, $14.9\%$, and $60.1\%$ for pressure, velocity, and TKE, respectively, demonstrating that excessive suppression of the latent velocity can remove the transport capacity required to reach the target distribution.

% (2) Subchannel
\subsection*{PWR subchannel}

\begin{figure}[h]
    \centering
    \includegraphics[width=0.99\textwidth]{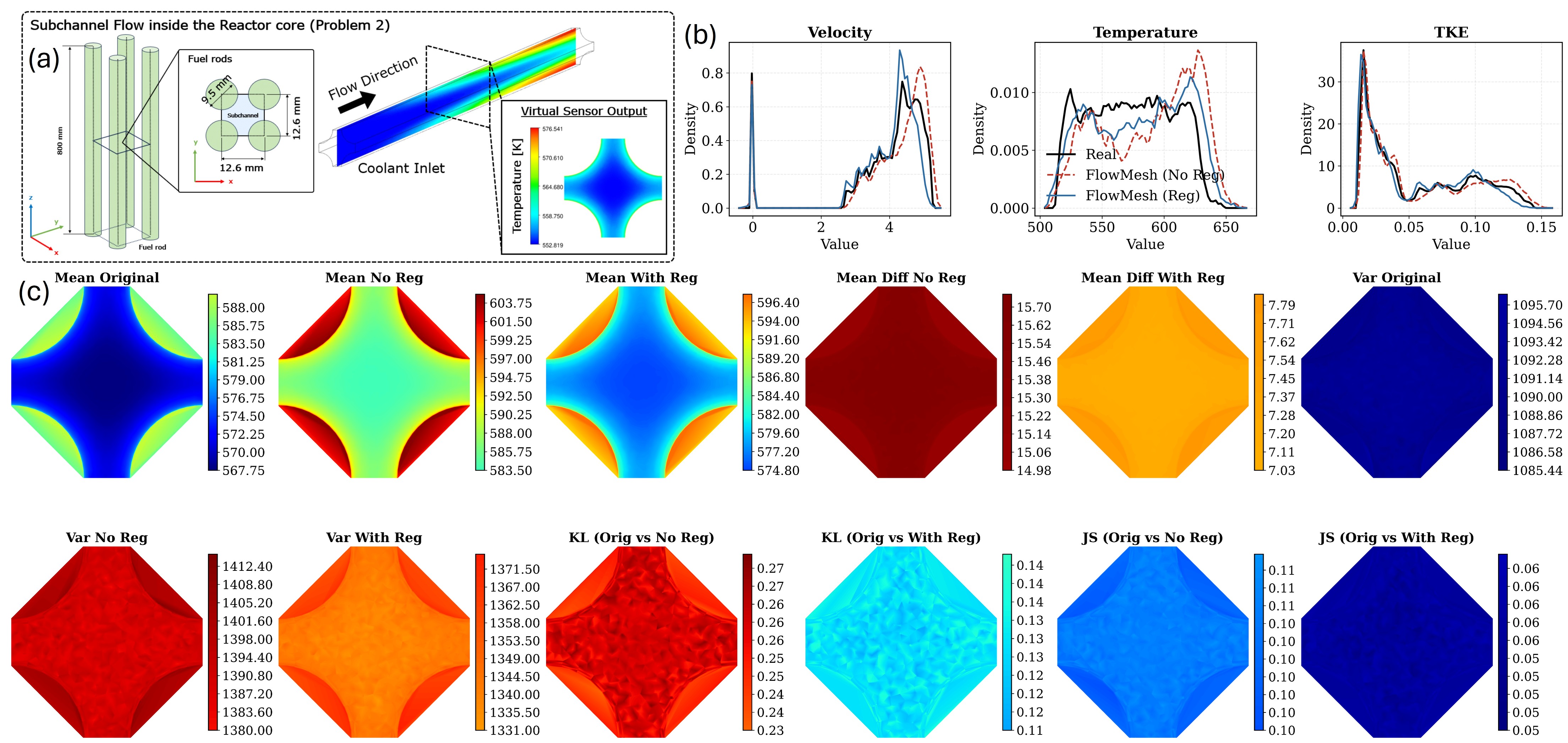}
    \caption{
    Pressurized water reactor (PWR) subchannel benchmark results on an unstructured mesh.
    \textbf{(a)} Schematic of the reactor subchannel geometry showing the fuel rods, coolant inlet conditions, flow direction, and extracted transverse cross-section, illustrated using the temperature field.
    \textbf{(b)} Probability density function (PDF) comparisons between the reference computational fluid dynamics (CFD) data and FlowMesh-generated samples for velocity magnitude, temperature, and turbulent kinetic energy (TKE), with and without regularization.
    \textbf{(c)} Node-wise diagnostics for temperature, including mean and variance fields from the reference data and generated samples with and without regularization, mean-error maps relative to the reference, and Kullback-Leibler (KL) and Jensen-Shannon (JS) divergence maps between the reference and generated node-wise distributions.
    }
    \label{fig:subchannel_case}
\end{figure}

We next test FlowMesh in a nuclear thermal-hydraulics setting in which distributional field modeling has direct engineering relevance. A pressurized-water-reactor (PWR) subchannel confines coolant between heated fuel rods, producing strongly nonuniform velocity, turbulence, and temperature fields on an irregular cross-section. Ensembles of such fields are relevant to uncertainty propagation, virtual sensing, and probabilistic characterization of thermal margins, but obtaining them through repeated CFD calculations is computationally expensive. The benchmark therefore tests whether latent generative transport remains accurate when the field distribution contains both coherent coolant motion and steep wall-adjacent thermal gradients. The operating variables comprise an axially varying wall heat-flux profile,

\begin{equation}
q(z) = A \sin\!\left( \frac{\pi z}{H} \right), \quad z \in [0,H],
\end{equation}

where $H$ is the heated length and $A$ is the heat-flux amplitude, together with the inlet temperature $T_{\text{in}} \in \mathbb{R}$ and inlet velocity magnitude $v_{\text{in}} \in \mathbb{R}$. The resulting input space is

\begin{equation}
\mathcal{U}_{\text{sub}} = \left\{ (q, T_{\text{in}}, v_{\text{in}}) \right\},
\end{equation}
where $q \in L^2([0,H])$ and $T_{\text{in}}, v_{\text{in}} \in \mathbb{R}$.

The response is recorded on a transverse plane $\mathcal{Y}_{\text{sub}} \subset \mathbb{R}^2$ at a fixed axial location $z=z_0$. At every mesh node, the model generates the velocity magnitude $\|\boldsymbol{v}(x,y)\|$, temperature $T(x,y)$, and turbulent kinetic energy $k(x,y)$. The output space therefore, is three-channel:

\begin{equation}
\mathcal{S}_{\text{sub}} = \big( L^2(\mathcal{Y}_{\text{sub}}) \big)^3,
\end{equation}

without interpolation of the unstructured field onto a Cartesian grid. We generated $5{,}000$ reference simulations in \textsc{ANSYS Fluent} on a three-dimensional unstructured mesh and extracted a common transverse section containing $1{,}733$ nodes. The operating parameters were sampled independently as

\begin{equation}
A \sim \mathcal{U}(540,660)\; \text{[kW/m$^2$]}, \quad 
T_{\text{in}} \sim \mathcal{U}(536.4,655.6)\; \text{[K]}, \quad 
v_{\text{in}} \sim \mathcal{U}(4.05,4.95)\; \text{[m/s]}.
\end{equation}

Each condition thus maps to an array in $\mathbb{R}^{1733\times3}$. This fixed but geometrically irregular node set isolates the distribution-learning problem from mesh interpolation error while retaining the rod-wall boundary layers that make the case challenging.

Across six fixed, matched training seeds, unregularized FlowMesh attains PDF coverages of $89.6\pm5.0\%$ for velocity, $81.6\pm6.6\%$ for temperature, and $92.5\pm3.8\%$ for TKE (Table~\ref{tab:subchannel_seed_metrics}). At $\lambda=10^{-3}$, the corresponding means increase to $92.5\pm2.5\%$, $83.7\pm12.1\%$, and $95.1\pm1.9\%$. The gain is most reproducible for velocity and TKE, for which the across-seed variability also decreases. Temperature is more sensitive to training seed at this weight. Increasing the regularization strength to $\lambda=10^{-1}$ reduces the mean temperature KL divergence from $0.1517$ to $0.0635$ and increases temperature PDF coverage from $81.6\%$ to $87.9\%$. This setting also gives the largest macro-averaged coverage, increasing it from $87.90\%$ without regularization to $91.67\%$. The six-seed experiment therefore supports a field-dependent effect of velocity regularization: moderate regularization improves the mean distributional fidelity of all three PWR fields, but the weight that is most favorable differs across channels and the temperature response is less seed-stable than the velocity and TKE responses.

The field statistics in Table~\ref{tab:Subchannel_model_stats} provide a complementary view of these distributional improvements. The $\lambda=10^{-1}$ model is closer to the original data in five of the nine reported statistics, primarily the field extrema, whereas the unregularized model is closer for all three field means and the TKE minimum. Thus, the improvements in KL divergence, JS divergence, and PDF coverage do not translate into uniformly improved means and extrema.

The spatial diagnostics in Fig.~\ref{fig:subchannel_case} complement these marginal metrics. The mean fields are reproduced across most of the coolant domain, while the largest KL and JS discrepancies remain concentrated near the heated rod surfaces, where the temperature and velocity gradients are steepest. This localization is consistent with temperature remaining the most challenging field, even though stronger regularization substantially improves its aggregate distributional metrics.

% ---------------------------------------------------------------------
% TABLE 4 -- PWR distributional metrics and AE diagnostic
% ---------------------------------------------------------------------

\begin{table*}[t]
  \centering
  \begin{threeparttable}
    \caption{PWR sub-channel distributional metrics over six fixed, matched
    training seeds (mean $\pm$ standard deviation). All configurations share
    the same architecture, data, and 100-epoch training schedule. FlowMesh is
    evaluated against autoencoder-reconstructed reference fields. Lower KL and JS and higher PDF coverage are better.}
    \label{tab:subchannel_seed_metrics}
    \vspace{4pt}

    \small
    \setlength{\tabcolsep}{3.5pt}
    \renewcommand{\arraystretch}{1.08}

    \begin{tabular}{@{}llccc@{}}
      \toprule
      \textbf{Model}
      & \textbf{Field}
      & \textbf{KL}~$\downarrow$
      & \textbf{JS}~$\downarrow$
      & \textbf{PDF Coverage (\%)}~$\uparrow$ \\
      \midrule

      \multirow{3}{*}{%
        \shortstack[l]{FlowMesh $\ell_1$\\No Reg.}}
      & Velocity
      & $0.0613 \pm 0.0454$
      & $0.0221 \pm 0.0169$
      & $89.6 \pm 5.0$ \\

      & Temperature
      & $0.1517 \pm 0.1041$
      & $0.0545 \pm 0.0339$
      & $81.6 \pm 6.6$ \\

      & TKE
      & $0.0350 \pm 0.0323$
      & $0.0118 \pm 0.0105$
      & $92.5 \pm 3.8$ \\
      \midrule

      \multirow{3}{*}{%
        \shortstack[l]{FlowMesh $\ell_1$, Reg.\\($\lambda=10^{-3}$)}}
      & Velocity
      & $\bm{0.0323 \pm 0.0186}$
      & $\bm{0.0114 \pm 0.0055}$
      & $\bm{92.5 \pm 2.5}$ \\

      & Temperature
      & $0.1350 \pm 0.1675$
      & $0.0536 \pm 0.0657$
      & $83.7 \pm 12.1$ \\

      & TKE
      & $\bm{0.0142 \pm 0.0095}$
      & $\bm{0.0050 \pm 0.0031}$
      & $\bm{95.1 \pm 1.9}$ \\
      \midrule

      \multirow{3}{*}{%
        \shortstack[l]{FlowMesh $\ell_1$, Reg.\\($\lambda=10^{-1}$)}}
      & Velocity
      & $0.0397 \pm 0.0352$
      & $0.0126 \pm 0.0100$
      & $92.4 \pm 3.3$ \\

      & Temperature
      & $\bm{0.0635 \pm 0.0286}$
      & $\bm{0.0242 \pm 0.0103}$
      & $\bm{87.9 \pm 4.0}$ \\

      & TKE
      & $0.0188 \pm 0.0179$
      & $0.0063 \pm 0.0057$
      & $94.7 \pm 2.5$ \\
      \bottomrule
    \end{tabular}
  \end{threeparttable}
\end{table*}

% ---------------------------------------------------------------------
% TABLE 6 -- PWR field statistics
% ---------------------------------------------------------------------
\begin{table*}[t]
  \centering
  \begin{threeparttable}
    \caption{PWR sub-channel field statistics. Each FlowMesh entry is the average of the corresponding statistic over the  6 training seeds. The Original row reports statistics from the reference dataset. Negative velocity-magnitude \& TKE values are physically inadmissible; the table reports raw model outputs prior to any non-negativity post-processing.}
    \label{tab:Subchannel_model_stats}
    \vspace{4pt}

    \begin{tabular}{
      l
      l
      S[table-format=3.3]
      S[table-format=-3.3]
      S[table-format=3.3]
    }
      \toprule
      \textbf{Data Origin}
      & \textbf{Field}
      & {\textbf{Mean}}
      & {\textbf{Min Value}}
      & {\textbf{Max Value}} \\
      \midrule

      \multirow{3}{*}{Original}
      & Velocity
      & 3.992 & 0.000 & 5.250 \\

      & Temperature
      & 573.880 & 512.260 & 649.680 \\

      & TKE
      & 0.056 & 0.011 & 0.146 \\
      \midrule

      \multirow{3}{*}{FlowMesh, No Reg.}
      & Velocity
      & 4.024 & -0.263 & 5.461 \\

      & Temperature
      & 575.795 & 503.519 & 659.990 \\

      & TKE
      & 0.057 & 0.009 & 0.156 \\
      \midrule

      \multirow{3}{*}{%
        FlowMesh, With Reg.\ ($\lambda=10^{-1}$)}
      & Velocity
      & 3.931 & -0.246 & 5.415 \\

      & Temperature
      & 576.075 & 503.580 & 658.834 \\

      & TKE
      & 0.054 & 0.008 & 0.153 \\
      \bottomrule
    \end{tabular}
  \end{threeparttable}
\end{table*}

% =====================================================================
% Table 5 (PWR) -- L1/L2 regularization comparison + autoencoder ceiling.
% AE label fix: the "AE reconstruction / diagnostic" label is split across the
% first two shaded rows instead of a \multirow. Reason: \multirow text is
% vertically centred (row 2) and the \rowcolor{gray!12} on rows 2-3 is painted
% on top of it, hiding the label. Splitting the label (no \multirow) avoids the
% overpaint entirely, so it always renders on the shaded background.
% =====================================================================
\begin{table}[H]
\centering
\caption{PWR sub-channel single-run comparison of $\ell_1$ and $\ell_2$ flow-matching supervision and the autoencoder reconstruction ceiling. For each loss, the unregularized baseline ($\lambda=0$) is compared with the selected regularized setting ($\lambda=10^{-3}$). The autoencoder reconstruction diagnostic passes $N=4,000$ reference fields through $D_\psi \circ E_\phi$ and evaluates them using the same metrics; it upper-bounds the distributional fidelity attainable by the latent flow model with this autoencoder. Lower KL, JS, and Wasserstein distance and higher PDF coverage are better.}
\label{tab:subchannel-ae-floor}
\small
\begin{tabular}{l l c c c c}
\toprule
\textbf{Model} & \textbf{Field} & \textbf{KL}~$\downarrow$ & \textbf{JS}~$\downarrow$ & \textbf{Wasserstein}~$\downarrow$ & \textbf{PDF Coverage (\%)}~$\uparrow$ \\
\midrule
\multirow{3}{*}{Flow Match -- No Reg ($\ell_1$, $\lambda{=}0$)}
  & Velocity             & $0.0276$ & $0.0112$ & $0.0650$ & $92.93$ \\
  & Temperature          & $0.0322$ & $0.0137$ & $1.8165$ & $92.77$ \\
  & TKE & $0.0117$ & $0.0044$ & $0.0016$ & $95.27$ \\
\midrule
\multirow{3}{*}{Flow Match -- With Reg ($\ell_1$, $\lambda^{\star}{=}10^{-3}$)}
  & Velocity             & $0.0175$ & $0.0075$ & $0.0255$ & $\mathbf{94.40}$ \\
  & Temperature          & $0.0310$ & $0.0134$ & $2.6921$ & $\mathbf{93.07}$ \\
  & TKE & $0.0047$ & $0.0018$ & $0.0006$ & $\mathbf{96.98}$ \\
\midrule
\multirow{3}{*}{Flow Match -- No Reg ($\ell_2$, $\lambda{=}0$)}
  & Velocity             & $0.0977$ & $0.0311$ & $0.1433$ & $85.10$ \\
  & Temperature          & $0.0275$ & $0.0105$ & $1.3739$ & $92.17$ \\
  & TKE & $0.0469$ & $0.0157$ & $0.0036$ & $89.77$ \\
\midrule
\multirow{3}{*}{Flow Match -- With Reg ($\ell_2$, $\lambda^{\star}{=}10^{-3}$)}
  & Velocity             & $0.0137$ & $0.0053$ & $0.0162$ & $94.30$ \\
  & Temperature          & $0.0282$ & $0.0109$ & $2.0575$ & $92.39$ \\
  & TKE & $0.0054$ & $0.0020$ & $0.0003$ & $96.25$ \\
\midrule
\rowcolor{gray!12}
\textbf{AE reconstruction} & Velocity    & $1.13\!\times\!10^{-4}$ & $4.2\!\times\!10^{-5}$ & $0.0044$ & $\mathbf{99.50}$ \\
\rowcolor{gray!12}
\textbf{(diagnostic)}      & Temperature & $8.8\!\times\!10^{-5}$  & $3.2\!\times\!10^{-5}$ & $0.0400$ & $\mathbf{99.60}$ \\
\rowcolor{gray!12}
                           & TKE         & $2.36\!\times\!10^{-4}$ & $7.9\!\times\!10^{-5}$ & $8.4\!\times\!10^{-5}$ & $\mathbf{99.43}$ \\
\bottomrule
\end{tabular}
\end{table}

% The separate $\ell_1$/$\ell_2$ sweep in Table~\ref{tab:subchannel-ae-floor} reaches its best joint result with $\ell_2$ supervision and $\lambda=10^{-3}$: PDF coverage is $94.30\%$ for velocity, $92.39\%$ for temperature, and $96.25\%$ for TKE. Because this sweep uses a different seed and training configuration, these values should not be compared row-by-row with the six-seed averages above. Instead, the experiment shows that the regularizer can improve all three channels under a suitable loss--weight combination, but that this benefit is not robust enough to claim a universal temperature improvement. The autoencoder-only evaluation attains more than $99.4\%$ coverage and KL below $2.4\times10^{-4}$ for every channel. The remaining temperature KL gap is therefore approximately $0.028$, indicating that the dominant error arises in latent transport rather than mesh reconstruction.

The separate single-run comparison in Table~\ref{tab:subchannel-ae-floor} attains its highest joint PDF coverage with $\ell_1$ supervision and $\lambda=10^{-3}$. This configuration achieves coverages of $94.40\%$ for velocity, $93.07\%$ for temperature, and $96.98\%$ for TKE, corresponding to a macro-average of $94.82\%$, compared with $94.31\%$ for the regularized $\ell_2$ configuration. Relative to the unregularized $\ell_1$ baseline, regularization increases coverage by $1.47$, $0.30$, and $1.71$ percentage points for velocity, temperature, and TKE, respectively. 

The temperature result is nevertheless metric-dependent. Under $\ell_1$ supervision, regularization slightly improves temperature coverage and KL divergence but increases the Wasserstein distance from $1.8165$ to $2.6921$. Under $\ell_2$ supervision, temperature coverage increases by only $0.22$ percentage points, while KL, JS, and Wasserstein distance all increase slightly. The single-run comparison therefore supports $\ell_1$ regularization as the best joint-coverage configuration and shows clearer benefits for velocity and TKE than for temperature. The autoencoder reconstruction diagnostic attains more than $99.4\%$ coverage and KL divergence below $2.4\times10^{-4}$ for every channel. Relative to this ceiling, the selected regularized $\ell_1$ configuration retains a temperature KL gap of approximately $0.0309$, indicating that most of the remaining distributional error is introduced by the latent flow stage rather than by autoencoder reconstruction.

Marginal distributional agreement does not imply that each generated realization satisfies the governing constraints. We therefore examined two complementary diagnostics for the regularized $\ell_1$ model with $\lambda=10^{-3}$. The global energy diagnostic is comparatively insensitive to the small negative TKE excursions: clipping these values to zero changes the system energy by less than $0.01\%$. The wall-boundary diagnostic exposes a more consequential limitation. Generated velocity at stationary walls has a mean residual of approximately $1.5$~m/s, and forcing the wall values to zero after generation removes approximately $5\%$ of the domain momentum and $3.5\%$ of its kinetic energy. Distributional fidelity and physical admissibility are therefore distinct objectives in the present model. The latter cannot be recovered through benign post-processing and will require hard boundary constraints or constraint-preserving parameterizations in the decoder or latent dynamics.

\subsection*{Heat Exchanger}

\begin{figure}[h]
    \centering
    \includegraphics[width=1\textwidth]{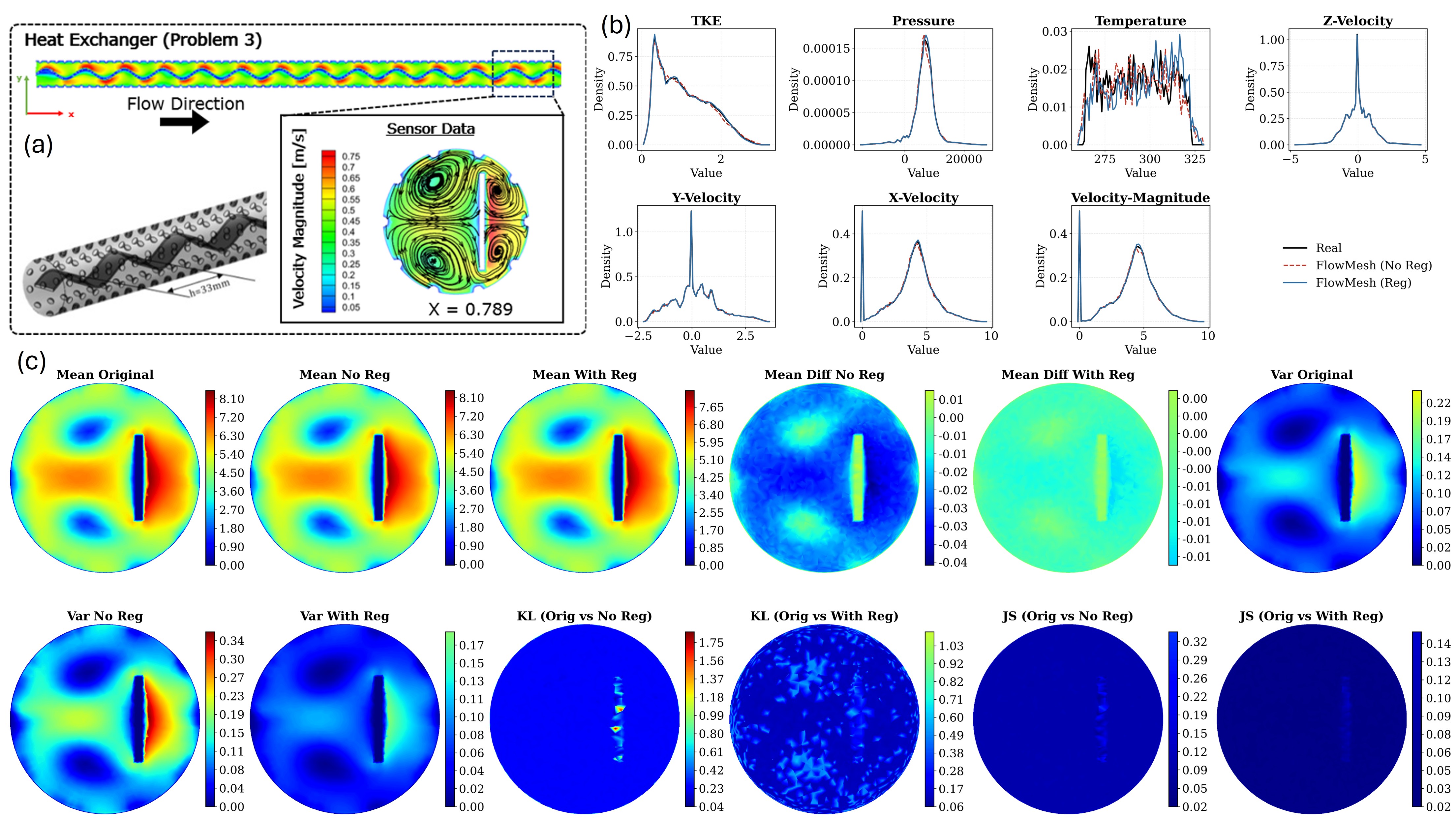}
    \caption{
    Heat exchanger benchmark results on an unstructured mesh.
    \textbf{(a)} Schematic of the heat exchanger configuration showing the flow direction, internal geometry, and transverse cross-section used for virtual sensing, with velocity magnitude visualized on the sensor plane.
    \textbf{(b)} Probability density function (PDF) comparisons between the reference computational fluid dynamics (CFD) data and FlowMesh-generated samples with and without regularization for turbulent kinetic energy (TKE), pressure, temperature, three velocity components, and velocity magnitude.
    \textbf{(c)} Node-wise diagnostics for velocity magnitude, including mean and variance fields from the reference data and generated samples with and without regularization, mean-difference maps relative to the reference, and Kullback--Leibler (KL) and Jensen--Shannon (JS) divergence maps between the reference and generated node-wise distributions.
    }
    \label{fig:hx_case}
\end{figure}

We next test FlowMesh on a heat-exchanger geometry to determine whether the trends observed in the PWR subchannel persist in a second energy-system configuration with different wall geometry and a larger set of thermofluid variables. The wall heat flux follows the same axial profile used in the PWR benchmark,
\begin{equation}
q(z) = A \sin\!\left( \frac{\pi z}{H} \right), \quad z \in [0,H],
\end{equation}
where $A$ is the heat-flux amplitude. The profile is applied uniformly along the streamwise $x$-direction,
\begin{equation}
q_{\text{wall}}(x,z) := q(z), \quad x \in [0,L],
\end{equation}
and the operating envelope additionally varies the inlet temperature $T_{\text{in}}$ and inlet velocity magnitude $u_{\text{in}}$.

The parameter space is therefore
\begin{equation}
\mathcal{U}_{\text{hx}} =
\left\{
q_{\text{wall}}(x,z)=A\sin\!\left(\tfrac{\pi z}{H}\right),
\;
T_{\text{in}}\in\mathbb{R},
\;
u_{\text{in}}\in\mathbb{R}
\right\}.
\end{equation}

The response is evaluated on a vertical cross-section $\mathcal{Y}_{\text{hx}}\subset\mathbb{R}^2$ at fixed transverse location $x=x_0$. The reported fields comprise turbulent kinetic energy, pressure, temperature, the three velocity components, and velocity magnitude:
\begin{equation}
\mathbf{s}_{\text{hx}}(z,y)
=
\left[
k,\,
p,\,
T,\,
u_z,\,
u_y,\,
u_x,\,
\|\mathbf{u}\|
\right]^\top ,
\qquad
(z,y)\in\mathcal{Y}_{\text{hx}}.
\end{equation}

These seven reported channels are generated jointly as separate outputs of the decoder, including $\|\mathbf u\|$, which is not computed from the generated components; they are the quantities used in the distributional comparisons of Tables~\ref{tab:HX_model_metrics} and~\ref{tab:HX_model_stats}.

% ---------------------------------------------------------------------
% HX metrics  -- tab:HX_model_metrics  (KL / JS / PDF coverage, 7 channels)
% ---------------------------------------------------------------------
\begin{table*}[t]
  \centering
  \begin{threeparttable}
    \caption{Comparison of FlowMesh metrics (KL, JS, PDF coverage) on the Heat-Exchanger dataset, averaged over \textbf{6 independent training seeds}. Coverage is evaluated against the autoencoder-reconstructed reference. Lower is better for KL and JS.}
    \label{tab:HX_model_metrics}
    \vspace{4pt}
    \begin{tabular}{ l l S[table-format=1.5] S[table-format=1.5] S[table-format=2.3] }
      \toprule
      \textbf{Model} & \textbf{Field} & {\textbf{KL}~$\downarrow$} & {\textbf{JS}~$\downarrow$} & {\textbf{PDF Coverage (\%)}~$\uparrow$} \\
      \midrule
      \multirow{7}{*}{FlowMesh No Regularizer}
        & TKE & 0.00382 & 0.00142 & 97.800 \\
        & Pressure                  & 0.00489 & 0.00177 & 96.162 \\
        & Temperature               & 0.18662 & 0.06622 & 80.288 \\
        & Z-Velocity                & 0.00038 & 0.00014 & 99.196 \\
        & Y-Velocity                & 0.00189 & 0.00070 & 98.586 \\
        & X-Velocity                & 0.00339 & 0.00134 & 97.784 \\
        & Velocity Magnitude        & 0.00321 & 0.00125 & 97.914 \\
      \midrule
      \multirow{7}{*}{FlowMesh With Regularizer ($\lambda{=}10^{-1}$)}
        & TKE & 0.00178 & 0.00063 & 98.426 \\
        & Pressure                  & 0.00260 & 0.00093 & 97.140 \\
        & Temperature               & 0.11142 & 0.04406 & 84.487 \\
        & Z-Velocity                & 0.00021 & 0.00008 & 99.393 \\
        & Y-Velocity                & 0.00083 & 0.00029 & 98.980 \\
        & X-Velocity                & 0.00112 & 0.00042 & 98.329 \\
        & Velocity Magnitude        & 0.00114 & 0.00043 & 98.398 \\
      \bottomrule
    \end{tabular}
  \end{threeparttable}
\end{table*}

% ---------------------------------------------------------------------
% HX stats  -- tab:HX_model_stats  (mean / min / max, 7 channels)
% ---------------------------------------------------------------------
\begin{table*}[t]
  \centering
  \begin{threeparttable}
    \caption{Comparison of statistical values of data across various fields (Heat-Exchanger). FlowMesh rows are the mean over \textbf{6 independent training seeds}; the Original row is as reported for the original dataset.}
    \label{tab:HX_model_stats}
    \vspace{4pt}
    \begin{tabular}{ l l S[table-format=4.4] S[table-format=-5.4] S[table-format=5.4] }
      \toprule
      \textbf{Data Origin} & \textbf{Field} & {\textbf{Mean}} & {\textbf{Min Value}} & {\textbf{Max Value}} \\
      \midrule
      \multirow{7}{*}{Original}
        & TKE & 1.0780 & 0.0490 & 2.9885 \\
        & Pressure                  & 6645.9000 & -13974.3800 & 25704.1000 \\
        & Temperature               & 292.3600 & 263.8500 & 322.5100 \\
        & Z-Velocity                & 0.0009 & -4.4910 & 4.5110 \\
        & Y-Velocity                & 0.1602 & -2.1800 & 3.4880 \\
        & X-Velocity                & 3.9170 & 0.0000 & 9.2520 \\
        & Velocity Magnitude        & 4.1920 & 0.0000 & 9.2730 \\
      \midrule
      \multirow{7}{*}{FlowMesh No Regularizer}
        & TKE & 1.0800 & 0.0303 & 3.1224 \\
        & Pressure                  & 6654.6775 & -14611.2389 & 26844.0859 \\
        & Temperature               & 293.0365 & 258.6028 & 328.3941 \\
        & Z-Velocity                & 0.0009 & -4.6226 & 4.6316 \\
        & Y-Velocity                & 0.1604 & -2.2456 & 3.5759 \\
        & X-Velocity                & 3.9197 & -0.1500 & 9.4946 \\
        & Velocity Magnitude        & 4.1951 & -0.1521 & 9.5210 \\
      \midrule
      \multirow{7}{*}{FlowMesh With Regularizer ($\lambda{=}10^{-1}$)}
        & TKE & 1.0732 & 0.0384 & 3.0997 \\
        & Pressure                  & 6614.1274 & -14477.3594 & 26646.2246 \\
        & Temperature               & 292.8551 & 258.5447 & 328.7100 \\
        & Z-Velocity                & 0.0009 & -4.5966 & 4.6061 \\
        & Y-Velocity                & 0.1599 & -2.2326 & 3.5566 \\
        & X-Velocity                & 3.9070 & -0.1323 & 9.4524 \\
        & Velocity Magnitude        & 4.1816 & -0.1296 & 9.4803 \\
      \bottomrule
    \end{tabular}
  \end{threeparttable}
\end{table*}

Table~\ref{tab:HX_model_metrics} shows that latent-velocity regularization improves the mean KL divergence, JS divergence, and PDF coverage for every reported heat-exchanger field. The macro-average PDF coverage increases from $95.390\%$ without regularization to $96.450\%$ at $\lambda=10^{-1}$, corresponding to a gain of $1.060$ percentage points. The largest improvement occurs for temperature, whose coverage increases from $80.288\%$ to $84.487\%$, while its KL divergence decreases from $0.18662$ to $0.11142$. The remaining fields already attain coverage above $96\%$ without regularization and consequently show smaller coverage gains, although their KL and JS divergences consistently decrease. These results indicate that regularization provides broad improvements across the heat-exchanger fields, with its principal benefit concentrated in the more challenging temperature distribution.

The field statistics in Table~\ref{tab:HX_model_stats} provide a complementary assessment. The regularized model is closer to the original data in 13 of the 21 reported mean, minimum, and maximum statistics, whereas the unregularized model is closer in seven and the two models tie in one. The regularized model improves most of the field extrema, while the unregularized model remains closer for several field means. Thus, the distributional improvements are broadly reflected in the summary statistics but are not uniform across every statistic. Because Tables~\ref{tab:HX_model_metrics} and~\ref{tab:HX_model_stats} report averages without seed-level standard deviations, these comparisons establish improvements in the mean metrics but do not by themselves demonstrate statistical significance across seeds.

\section*{Discussion}

\begin{figure}[t]
    \centering
    \includegraphics[width=\linewidth]{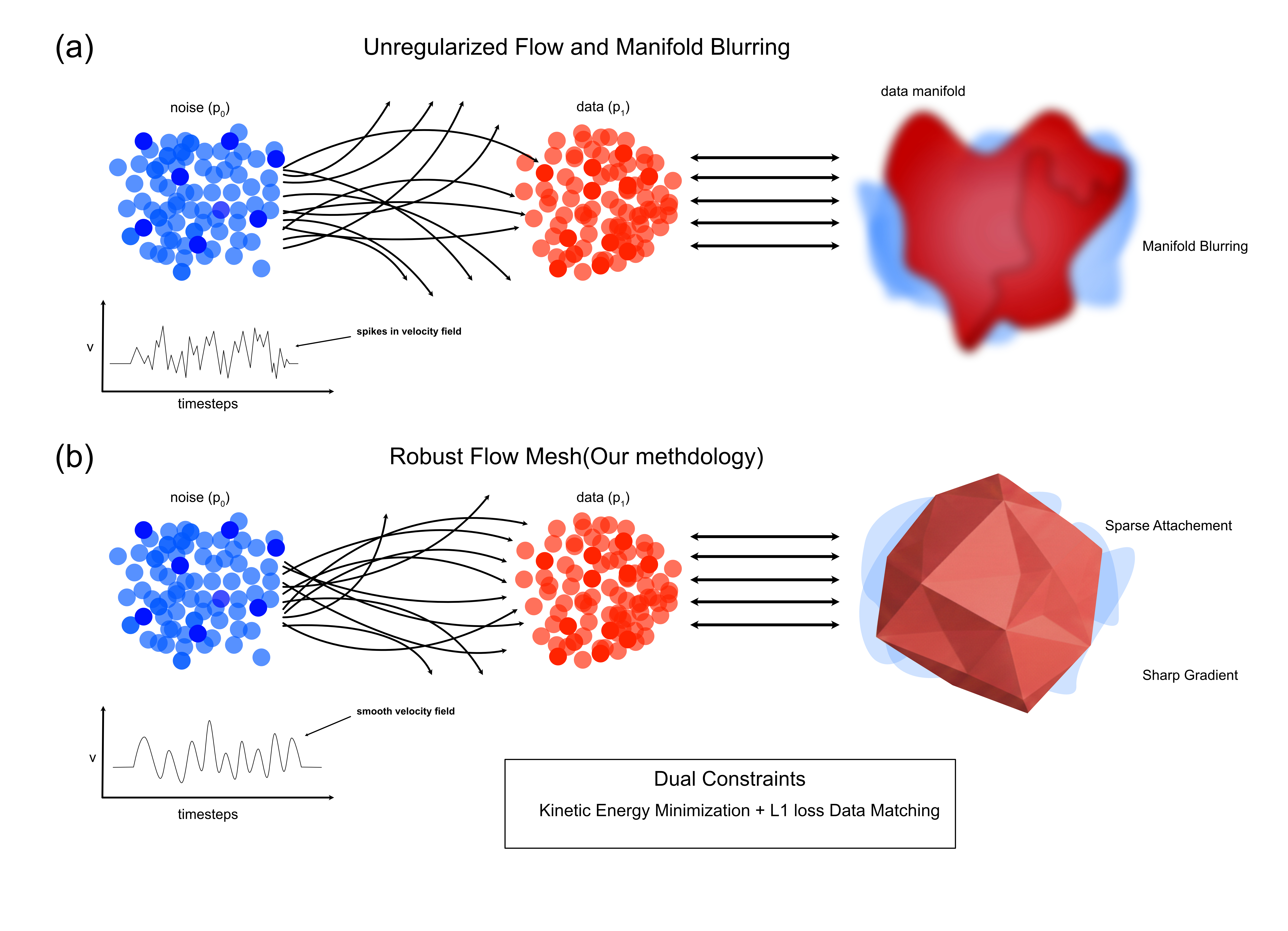}
   \caption{\textbf{Conceptual interpretation of latent-velocity regularization.}
\textbf{(a)} Unregularized flow matching is free to use latent velocities of any magnitude required by the data-matching objective. Large or rapidly varying transport fields can increase sensitivity of the numerical trajectory to local approximation error. The illustration is schematic and should not be interpreted as a measured trajectory or direct evidence of ``manifold detachment.''
\textbf{(b)} FlowMesh augments the flow-matching loss with an $\ell_1$ penalty on the latent velocity magnitude. The regularization sweep in Fig.~\ref{fig:lambda_sweep} shows that increasing $\lambda$ is empirically associated with a reduction in the sampled Jacobian norm of the learned velocity field, while moderate values can improve distributional fidelity. The implemented penalty is not a Dirichlet-energy objective and does not establish Wasserstein-geodesic optimality or formal contractivity.}
    \label{fig:velocity_regularization}
\end{figure}

\begin{table*}[t]
\centering
\caption{Comparison of PDF coverage (\%) between FlowMesh, vanilla flow
matching (Lipman et al.), and a latent DDPM baseline across benchmark
datasets. FlowMesh and vanilla flow matching values are unweighted
macro-averages across all reported fields using six fixed, matched training
seeds. FlowMesh uses $\ell_1$ velocity regularization with
$\lambda=10^{-1}$, whereas vanilla flow matching uses the identical model
with $\lambda=0$. All three models share the same PointNet autoencoder and
latent space. Higher values indicate better distributional fidelity.}
\label{tab:pdf_comparison}

\begin{tabular}{lccc}
\toprule
\textbf{Dataset}
& \textbf{FlowMesh (Ours)}
& \textbf{Flow Matching (Lipman et al.)}
& \textbf{Latent DDPM} \\
\midrule
Lid-driven Cavity
& \textbf{96.97}
& 96.47
& 85.39 \\

PWR Subchannel
& \textbf{91.67}
& 87.90
& 46.30 \\

Heat Exchanger
& \textbf{96.45}
& 95.39
& 82.70 \\
\bottomrule
\end{tabular}
\end{table*}

\paragraph{Deterministic transport versus latent diffusion}
Table~\ref{tab:pdf_comparison} provides the most direct comparison between the two generative mechanisms considered here. FlowMesh achieves higher macro-averaged PDF coverage than the latent DDPM on all three benchmarks, with differences of approximately $11.6$, $45.4$, and $13.8$ percentage points for the LDC, PWR subchannel, and heat exchanger, respectively. Because the two approaches use the same PointNet autoencoder and therefore operate on the same encoded data, the comparison controls for the principal geometric representation. It does not, however, constitute a universal comparison between flow matching and diffusion: the generative networks differ in parameterization and size, and the diffusion baseline uses a standard cosine-schedule DDPM with 250-step DDIM sampling rather than an exhaustively optimized scientific diffusion model. The experiments therefore establish that deterministic flow matching is substantially more effective than this architecture-controlled latent DDPM configuration on the present datasets.

The result is consistent with two plausible advantages of FlowMesh. First, its sampling trajectory is obtained by integrating a learned velocity field rather than repeatedly estimating a reverse denoising update. Second, the adaptive ODE solver requires fewer network evaluations than the 250-step DDIM sampler used here. These mechanisms are consistent with the measured accuracy and cost differences but are not independently isolated by the present experiments. We therefore interpret the diffusion comparison as evidence for the effectiveness of deterministic latent transport on these mesh-based physical-field distributions, not as evidence that stochastic generative modeling is intrinsically incompatible with physical systems.

\paragraph{Computational cost: deterministic transport versus iterative denoising}
Table~\ref{tab:compute} compares the measured computational cost of FlowMesh and the latent DDPM. Under the common 100-epoch training budget and batch size of 128, the recorded training times are $7.6$~min for FlowMesh and $39.5$~min for DDPM. This wall-clock difference should not be interpreted as an algorithm-only speed-up because the reported implementations use different numerical precision, bf16 for FlowMesh and fp32 for DDPM, and different network sizes. The sampling comparison is more directly interpretable. FlowMesh requires $94.7\pm11.6$ adaptive velocity-network evaluations per generated sample, whereas the selected DDIM sampler uses 250 denoising evaluations. The resulting measured inference costs are approximately $1.4$~TFLOP and $2.6$~TFLOP per sample, respectively.

End-to-end, one FlowMesh realization, including latent ODE integration and decoder reconstruction, requires $1.23\pm0.15$~s on a single NVIDIA A100 GPU, compared with $2.24$~s for the DDIM implementation. The larger practical gain is relative to high-fidelity simulation: depending on the benchmark, FlowMesh is $585$--$1{,}902$ times faster than the corresponding \textsc{ANSYS Fluent} calculation (Table~\ref{tab:inference_time}). Thus, in the implementations evaluated here, deterministic latent transport provides both higher distributional fidelity and lower sampling cost than the latent DDPM while reducing the cost of generating additional thermofluid realizations by approximately three orders of magnitude relative to CFD.

\begin{table}[h!]
\centering
\caption{Computational cost of FlowMesh versus the latent DDPM baseline.
Both models operate on identical PointNet latent spaces; reported differences reflect only the generative component.   Measured on one NVIDIA A100 (40 GB) GPU on the 4,096-dimensional latent (reshaped to $1\times64\times64$), which is identical across datasets; training times were measured on the PWR subchannel latents, with FlowMesh at $\lambda=0$; FLOPs from the PyTorch operator-level profiler. Training uses a matched 100-epoch budget and a batch size of ~128; FlowMesh trains in mixed precision (bf16), and DDPM in fp32, following each reference implementation.}
\label{tab:compute}
\begin{tabular}{lcc}
\toprule
\textbf{Metric} & \textbf{FlowMesh} & \textbf{Latent DDPM (shared PointNet encoder)} \\
\midrule
Network parameters               & 20.8\,M                    & 6.1\,M          \\
Forward pass                     & 15.1\,GFLOP                & 10.6\,GFLOP     \\
Training wall-clock (100 epochs) & 7.6\,min                   & 39.5\,min       \\
Peak training memory             & 9.4\,GB                    & 6.1\,GB         \\
\midrule
NFE per sample                   & $94.7 \pm 11.6$ (adaptive) & 250 (DDIM) \\
Inference cost per sample        & $\sim$1.4\,TFLOP           & $\sim$2.6\,TFLOP \\
Wall-clock per sample            & $1.23 \pm 0.15$\,s         & $2.24 \pm 0.003$\,s \\
Sampling memory (batch 1)        & 200\,MB                    & 310\,MB \\
Hardware                         & A100 (40\,GB)              & A100 (40\,GB)   \\
\bottomrule
\end{tabular}
\end{table}

% ---------------------------------------------------------------
% Computation Time Table — FlowMesh per dataset
% Place this after Table~\ref{tab:compute} or in the Methods section.
% Fill in [FILL: ...] cells with your measured values.
% ---------------------------------------------------------------

\begin{table}[h!]
  \centering
  \begin{threeparttable}
  \caption{Per-sample computation time for each benchmark dataset.CFD reference times correspond to single-case \textsc{ANSYS Fluent} runs on a 12-core Xeon CPU. FlowMesh inference is measured on a single NVIDIA A100 (40\,GB) GPU (batch size~1, adaptive Dormand--Prince ODE solver, regularised model $\lambda{=}10^{-3}$), averaged over 1{,}000 samples following 20 warm-up runs.Speed-up is the ratio of the CFD wall-clock time to the FlowMesh wall-clock time.}
  \label{tab:inference_time}
  \begin{tabular}{
    l
    S[table-format=5.0]
    S[table-format=4.0]
    S[table-format=1.2, table-figures-uncertainty=2]
    S[table-format=5.0]
  }
    \toprule
    \textbf{Dataset}
      & {\textbf{Mesh Nodes}}
      & {\textbf{CFD (s/sample)}}
      & {\textbf{FlowMesh (s/sample)}}
      & {\textbf{Speed-up$^\dagger$}} \\
    \midrule
    Lid-driven Cavity (LDC)  & 4225  & 2340 & 1.23 \pm 0.15 & 1902 \\
    PWR Subchannel           & 1733  &  720 & 1.23 \pm 0.15 & 585 \\
    Heat Exchanger (HX)      & 3977  & 1800 & 1.23 \pm 0.15 & 1463 \\
    \bottomrule
  \end{tabular}
  
  \end{threeparttable}
\end{table}

\paragraph{Relation to neural operators}
The present work targets distributional generation rather than deterministic forward regression. Neural operators such as DeepONet \cite{lu2021learning}, MIONet \cite{jin2022mionet}, S-DeepONet \cite{goswami2022physicsinformeddeepneuraloperator}, the Fourier neural operator \cite{li2020fourier}, geometry-informed operators \cite{li2023geometry}, and transformer-based operators are commonly trained to predict a field associated with a specified input function or parameter set. Under standard pointwise regression losses, these models are naturally evaluated through prediction error at that operating condition. FlowMesh instead generates an ensemble of fields whose distribution approximates the distribution represented by the training simulations. Because each reference solution is a deterministic RANS field, this distribution is the push-forward of the operating-condition distribution used to generate the dataset, $p(\mathbf s)=G_{\#}\,p(u)$ (Appendix~1.3). It therefore characterizes the sampled operating envelope (uniform over the ranges of Eq.~(10) for the PWR subchannel) rather than a physical operational-uncertainty distribution, and when $u$ and $p(u)$ are known the same ensemble can also be obtained by evaluating a deterministic operator on sampled inputs; FlowMesh learns it from field samples alone.

The distinction is particularly relevant in energy applications in which an ensemble, rather than a single best-estimate field, is required. Examples include estimating the distribution of peak coolant temperature across an uncertain operating envelope, generating synthetic field ensembles for downstream uncertainty propagation, and producing candidate states for sensor-placement or probabilistic-analysis studies. The present implementation should not, however, be interpreted as a conditional neural operator: as formulated in the Methods, the latent velocity field is not explicitly conditioned on a requested operating point. Conditional generation at prescribed heat flux, inlet temperature, or inlet velocity is therefore a natural extension rather than a capability demonstrated here.

Probabilistic neural operators, stochastic operator models, ensembles, and Bayesian formulations also address uncertainty in function-space prediction and are scientifically relevant comparators. We do not claim that generative flow matching is the only mechanism capable of representing field distributions. The comparison in this work is deliberately narrower: given a common PointNet representation, we compare deterministic latent flow matching, its velocity-regularized variants, and a standard latent DDPM. Deterministic neural operators remain complementary when fast point prediction at specified operating conditions is the primary objective.

\paragraph{Decomposing reconstruction and transport error}
The two-stage architecture introduces two distinct sources of error: compression and reconstruction through the autoencoder, and distributional transport in latent space. The autoencoder-only diagnostic in Table~\ref{tab:subchannel-ae-floor} attains PDF coverage above $99.4\%$ and KL divergence below $2.4\times10^{-4}$ for all three PWR fields. By contrast, the selected regularized $\ell_1$ flow model yields a temperature KL divergence of $0.0310$ and PDF coverage of $93.07\%$. The approximately $0.0309$ KL gap relative to the autoencoder reconstruction therefore arises predominantly after encoding, indicating that latent distribution transport, rather than mesh reconstruction, is the principal remaining source of error in this benchmark. We use the autoencoder result as an empirical reconstruction reference rather than as a formal mathematical upper bound on every possible decoded distribution.

\paragraph{Limitations}
Several limitations define the present scope. First, all experiments use a fixed mesh for each physical system. PointNet provides permutation invariance to node ordering, but the study does not demonstrate zero-shot transfer across mesh topologies, resolutions, or geometries. Second, the generative velocity field is not explicitly conditioned on operating parameters in the formulation reported here. FlowMesh therefore learns the field distribution over the sampled operating envelope rather than a conditional distribution at an arbitrary prescribed operating point. Extending the model to conditional transport is necessary for many digital-twin applications.

Third, distributional fidelity does not guarantee physical admissibility. The PWR model exhibits a mean no-slip wall residual of approximately $1.5$~m/s, and post-hoc correction changes domain momentum and kinetic energy non-negligibly. Hard boundary conditions, conservation constraints, or constraint-preserving output parameterizations will therefore be required before the method can be considered for safety-critical operational use. Small negative values also occur in quantities such as velocity magnitude and TKE, showing that positivity is not enforced by construction.

Fourth, the regularization weight is problem dependent. The principal six-seed comparisons use $\lambda=10^{-1}$, whereas the separate PWR loss ablation and computational-cost measurements use $\lambda=10^{-3}$. The sweep demonstrates a broad useful regime but does not provide an automatic selection rule. Fifth, the latent DDPM shares the same PointNet representation and training data but is not parameter-matched to the FlowMesh velocity network and uses a standard 250-step DDIM sampler. The reported diffusion gap therefore applies to this controlled baseline rather than to the full class of diffusion or score-based generative models.

Finally, the present benchmarks are fixed-geometry thermofluid field ensembles. Generalization to transient three-dimensional systems, discontinuous multiphase or reacting flows, previously unseen geometries, and out-of-distribution operating regimes remains to be established. These limitations bound the current conclusion to deterministic generative transport of distributions on the fixed unstructured geometries evaluated here.

The present results establish a narrower foundation for those applications. Permutation-invariant encoding can compress fixed unstructured-mesh fields with little reconstruction loss, deterministic flow matching can reproduce their distributions more accurately and cheaply than the latent DDPM evaluated here, and moderate velocity regularization can improve transport in the benchmarks for which the unregularized model is less accurate. The next scientific step is not simply to scale the architecture, but to integrate conditional operating variables, hard conservation and boundary constraints, topology-changing geometries, and transient dynamics into the generative process. Demonstrating those capabilities would determine whether deterministic latent transport can progress from efficient ensemble generation toward trustworthy generative digital twins for nuclear and other safety-critical energy systems.

\begin{figure}[t]
    \centering
    \includegraphics[width=\textwidth]{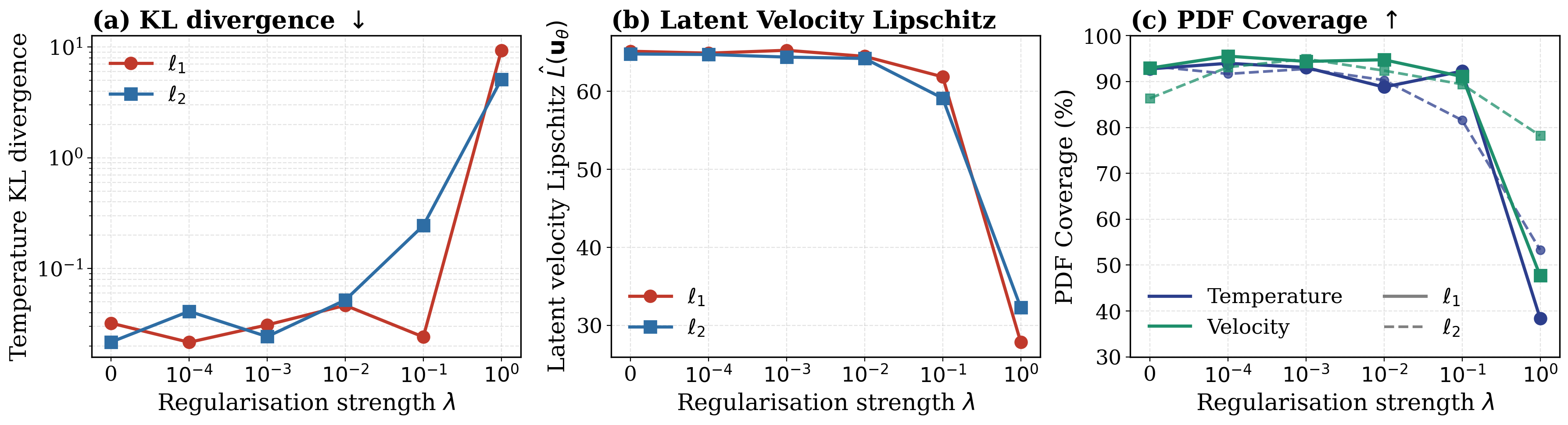}
\caption{\textbf{Regularization sweep maps the accuracy--stability trade-off in latent transport.}
PWR subchannel benchmark, single-run sweep.
\textbf{(a)} Temperature KL divergence as a function of regularization strength $\lambda\in\{0,10^{-4},10^{-3},10^{-2},10^{-1},1\}$. Under $\ell_1$ supervision, KL decreases from the unregularized value of approximately $0.032$ to approximately $0.022$ at $\lambda=10^{-4}$ and remains comparatively low through moderate $\lambda$, before increasing sharply at $\lambda=1$. The $\ell_2$ model deteriorates at smaller regularization weights.
\textbf{(b)} Empirical Jacobian-norm proxy
$\hat{J}=\max_{(\mathbf{z},t)\in\mathcal{S}}\|\nabla_{\mathbf{z}}\mathbf{u}_\theta(\mathbf{z},t)\|_F$
estimated on the sampled evaluation set $\mathcal{S}$. The proxy decreases from approximately $65$ at small $\lambda$ to approximately $28$ for $\ell_1$ and $32$ for $\ell_2$ at $\lambda=1$. This is an empirical measure of local velocity-field variation, not a proof of a global Lipschitz bound.
\textbf{(c)} Temperature and velocity PDF coverage versus $\lambda$. Small-to-moderate $\ell_1$ regularization improves both fields in this single run, whereas excessive regularization causes both to collapse. Together, the panels show that useful regularization occurs in an intermediate regime: reducing latent-velocity freedom can improve distributional transport, but suppressing it too strongly prevents the model from reaching the target distribution.}
    \label{fig:lambda_sweep}
\end{figure}

\section*{Conclusion}
FlowMesh shows that deterministic latent transport can generate distributions of high-dimensional physical fields on fixed unstructured meshes without interpolation to an auxiliary grid. Across lid-driven-cavity, PWR-subchannel, and heat-exchanger benchmarks, it attains 91.7–97.0\% macro-averaged PDF coverage and outperforms a latent DDPM built on the same PointNet representation by 11.6–45.4 percentage points, while reducing the cost of each additional realization by roughly three orders of magnitude relative to the reference CFD. The effect of velocity regularization is regime dependent: it leaves the already accurate cavity model essentially unchanged but improves mean coverage where unregularized transport is less accurate, most strongly for temperature. An autoencoder-only diagnostic locates the remaining error primarily in latent transport rather than in mesh reconstruction.

Nuclear thermal hydraulics provides both the most demanding test case and the clearest limitation. In the PWR subchannel, high distributional fidelity coexists with a mean no-slip residual of approximately 1.5 m/s at stationary walls, showing that matching field statistics does not by itself enforce physical admissibility. Integrating hard boundary and conservation constraints into the decoder or the latent dynamics is therefore the next scientific challenge, together with conditioning on operating variables and extending the approach to topology-changing geometries and transient dynamics.

With such constraints in place, inexpensive ensemble generation after a high-fidelity dataset has been constructed could support rapid uncertainty propagation across coolant states, synthetic field ensembles for sensor-placement studies, and probabilistic exploration of thermal margins in nuclear reactors, heat exchangers, and other energy systems in which complex geometry makes repeated CFD evaluation costly.

\section*{Methods}

\paragraph{Overview and rationale for the coupling}
FlowMesh separates geometric compression from distributional generation. A PointNet-based autoencoder first maps a field defined on a fixed unordered node set to a compact latent representation. Because the same node-wise transformation is followed by a symmetric pooling operation, the encoded representation is invariant to permutations of the input ordering. A flow-matching model is then trained on the encoded states and learns a deterministic time-dependent velocity field whose ordinary differential equation transports samples from a Gaussian reference distribution to the latent data distribution. The generative variable $t\in[0,1]$ parameterizes this probability-transport path and is not physical simulation time.

The autoencoder and flow model are trained sequentially. This design allows reconstruction error and latent-transport error to be diagnosed separately, as demonstrated by the autoencoder-only experiment in Table~\ref{tab:subchannel-ae-floor}. We additionally augment the flow objective with an $\ell_1$ penalty on the latent velocity magnitude. The penalty does not directly constrain derivatives of the velocity field and therefore is not a formal Lipschitz or Dirichlet-energy regularizer. Its influence on local velocity-field variation is assessed empirically through the Jacobian-norm diagnostic in Fig.~\ref{fig:lambda_sweep}. The regularization sweep is used to determine whether the penalty improves generation, leaves it unchanged, or suppresses transport excessively.

Operating in latent space also reduces the dimensionality presented to the generative model relative to the original mesh fields. The actual computational consequences of this compression are quantified through measured network evaluations, FLOPs, memory, and wall-clock sampling time in Table~\ref{tab:compute}, rather than inferred from dimensionality alone.

\subsection*{Latent Encoding of Unstructured Mesh}
Each sample is a field on a fixed mesh of $N$ nodes, $\mathcal X=\{(\mathbf x_i,\mathbf f_i)\}_{i=1}^N$, where $\mathbf x_i\in\mathbb R^2$ are node coordinates and $\mathbf f_i\in\mathbb R^d$ are the node-associated physical quantities ($d=3$ for the cavity and subchannel, $d=7$ for the heat exchanger). Each field channel is standardized to zero mean and unit variance over all samples and nodes; coordinates are left unscaled. A separate autoencoder is trained for each dataset.

The encoder $E_\phi$ follows PointNet~\cite{qi2017pointnet} but uses mean rather than max pooling. Each node input $\mathbf u_i=[\mathbf x_i;\mathbf f_i]\in\mathbb R^{2+d}$ passes through a shared node-wise MLP with widths 64, 128 and 256 and ReLU activations,
\begin{equation}
\mathbf h_i=\mathrm{MLP}_\phi(\mathbf u_i)\in\mathbb R^{256},
\end{equation}
and the node embeddings are averaged,
\begin{equation}
\mathbf g=\frac{1}{N}\sum_{i=1}^{N}\mathbf h_i ,
\end{equation}
which makes the encoding invariant to node ordering. A linear layer then maps the pooled feature to the latent,
\begin{equation}
\mathbf z=\operatorname{reshape}_{1\times 64\times 64}\!\left(\mathbf W\mathbf g+\mathbf b\right),\qquad \mathbf W\in\mathbb R^{4096\times 256}.
\end{equation}
The reshape to a $1\times64\times64$ array has no spatial meaning; it allows both generative models to use standard image U-Net architectures. Because $\mathbf z$ is an affine function of $\mathbf g$, the latent distribution lies on an affine subspace of dimension at most 256 within $\mathbb R^{4096}$. The latent size is identical for all three datasets, compared with $4225\times3$, $1733\times3$ and $3977\times7$ field values for the cavity, subchannel and heat exchanger, respectively.

The decoder $D_\psi$ maps $\mathbf z$ back to the fields on the fixed node set. Two convolutional layers ($1\rightarrow16\rightarrow32$ channels) are followed by adaptive average pooling to $8\times8$ and fully connected layers whose output has size $N\times d$, so that each output entry corresponds to a fixed node of the mesh. Node coordinates are encoder inputs only and are not reconstructed; each trained decoder is therefore specific to its mesh.

The autoencoder minimizes the mean-squared reconstruction error of the standardized fields,
\begin{equation}
\mathcal L_{\mathrm{AE}}=\frac{1}{Nd}\sum_{i=1}^{N}\bigl\|\hat{\mathbf f}_i-\mathbf f_i\bigr\|_2^2 ,
\end{equation}
using Adam (learning rate $10^{-4}$, batch size 32, 50 epochs). Reconstruction fidelity is assessed by the relative mean-squared error of the de-standardized fields and by the distributional diagnostic in Table~6.

\subsection*{Model architecture}
In our methodology, we have employed Flow Matching (FM), a simulation-free framework that learns a time-dependent velocity field whose flow transports samples from a simple source distribution to the target data distribution. FM constructs a continuous-time probability path $p_t$ interpolating between a Gaussian source distribution $p_0=\mathcal{N}(\mathbf{0},\mathbf{I})$ and the target latent-data distribution $p_1=p_{\mathrm{data}}$, and trains a neural velocity field $u_\theta(t,\mathbf{z})$ to reproduce the instantaneous velocities that generate this path. The flow defined by $u_\theta$ is then used in sampling by integrating with the following ODE:
\begin{equation}
    \frac{d}{dt} \psi_t(z)=u_{\theta}\big(t,\psi_t(z)\big) , 
    \quad
    \psi_0(z)=z
\end{equation}

\subsubsection*{Model parametrization}
The velocity field $\mathbf u_\theta(\mathbf z,t)$ is parameterized by a U-Net of the ADM architecture~\cite{dhariwal2021diffusion} operating on the $1\times64\times64$ latent array, with 64 base channels, channel multipliers $(1,2,3,4)$, two residual blocks per resolution, single-head self-attention at the $16\times16$ feature resolution, and no dropout (20.8~M parameters). The generative time $t$ enters through sinusoidal embeddings. The network receives no other input: neither the operating parameters nor any observation of the field is provided, so the model learns the unconditional latent distribution $p_1$. Its output is a velocity with the same shape as $\mathbf z$. The same configuration is used for all three datasets and is trained in bfloat16 mixed precision.

\subsubsection*{Training }
Using a straight-line interpolation between independent samples
$\mathbf{z}_0\sim p_0$ and $\mathbf{z}_1\sim p_1$, the network is
trained by minimizing either the $\ell_1$ or $\ell_2$ error between
the predicted and target velocities:

\begin{equation}
\mathcal{L}_{\mathrm{CFM}}^{\ell_1}(\theta)
=
\mathbb{E}_{t,\mathbf{z}_0,\mathbf{z}_1}
\left[
\left\|
\mathbf{u}_{\theta}(\mathbf{z}_t,t)
-
\mathbf{u}_t(\mathbf{z}_t\mid\mathbf{z}_1)
\right\|_1
\right],
\end{equation}

or

\begin{equation}
\mathcal{L}_{\mathrm{CFM}}^{\ell_2}(\theta)
=
\mathbb{E}_{t,\mathbf{z}_0,\mathbf{z}_1}
\left[
\left\|
\mathbf{u}_{\theta}(\mathbf{z}_t,t)
-
\mathbf{u}_t(\mathbf{z}_t\mid\mathbf{z}_1)
\right\|_2^2
\right].
\end{equation}

Here,
\begin{equation}
\mathbf{z}_t
=
(1-t)\mathbf{z}_0+t\mathbf{z}_1,
\qquad
\mathbf{u}_t(\mathbf{z}_t\mid\mathbf{z}_1)
=
\frac{\mathbf{z}_1-\mathbf{z}_t}{1-t}.
\end{equation}

Unless otherwise stated, results labeled $\ell_1$ use
$\mathcal{L}_{\mathrm{CFM}}^{\ell_1}$, whereas results labeled
$\ell_2$ use $\mathcal{L}_{\mathrm{CFM}}^{\ell_2}$. Following~\cite{lipman2024flow}, ``conditional'' in $\mathcal L_{\rm CFM}$ refers to conditioning the target velocity $\mathbf u_t$ on the endpoint $\mathbf z_1$, not on physical inputs.

Here, $z_t$ is the interpolated sample at time $t$, and $u_t$ is the ground-truth instantaneous velocity. During each training iteration, batches of $(z_0, z_1)$ pairs are drawn, and an independent $t$ value is sampled uniformly in $[0, 1]$ for each pair. The interpolated samples $z_t$ and their corresponding target velocities $u_t$ are computed analytically. The network parameters are optimized using the Adam optimizer ($\beta_{1} = 0.9$, $\beta_{2} = 0.999$) with a constant learning rate.

\subsubsection*{Sampling}

To generate novel samples from the learned model, we begin by drawing an initial point $z_0 \sim p_0$ from the Gaussian distribution and evolving it forward in continuous time according to the learned probability flow $\dot{z}_t = u_\theta(z_t, t)$. For numerical integration, we employ the adaptive Dormand-Prince method, which provides a favorable balance between accuracy and computational efficiency for this task. The solver adjusts its step size adaptively based on local error estimates, with relative and absolute tolerances tuned on a held-out validation set.

\subsubsection*{Latent diffusion baseline}

To separate geometric representation from the choice of generative mechanism, the diffusion baseline uses the same trained PointNet autoencoder as FlowMesh, including the same encoder $E_\phi$, decoder $D_\psi$, latent representation, and reconstruction objective $\mathcal{L}_{\mathrm{AE}}$. The encoded training samples presented to the two generative models are therefore identical. The latent diffusion model is a denoising diffusion probabilistic model (DDPM) \cite{ho2020denoising} trained with a cosine noise schedule and sampled with a 250-step DDIM procedure. The noise-prediction network is a U-Net operating on the same $1\times64\times64$ latent array, with encoder widths $(32,64,128,256)$, bottleneck widths $(256,256,128)$, 128-dimensional time embeddings, and four-head self-attention with group normalization in every block (6.1~M parameters), using $T=1000$ diffusion steps.

The comparison controls the geometric representation, dataset, batch size, and evaluation pipeline, but it is not parameter matched: the FlowMesh and DDPM generative networks have different parameter counts and computational graphs, as reported in Table~\ref{tab:compute}. Their training implementations also use different numerical precision. Consequently, differences in distributional accuracy isolate the generative pipelines under a shared latent representation, but should not be interpreted as a theorem that flow matching is intrinsically superior to all diffusion architectures.

\subsubsection*{Latent-velocity regularization and stability diagnostics}

Flow matching learns a velocity field
$\mathbf{u}_\theta(\mathbf{z},t)$ whose ODE
\begin{equation}
    \frac{d\mathbf{z}_t}{dt}
    =
    \mathbf{u}_\theta(\mathbf{z}_t,t),
    \qquad
    \mathbf{z}_0\sim p_0,
\end{equation}
transports probability mass from the reference distribution toward the latent data distribution. Under the usual regularity assumptions, the corresponding density satisfies the continuity equation
\begin{equation}
    \frac{\partial p_t}{\partial t}
    +
    \nabla_{\mathbf{z}}\cdot
    \left(
    p_t\,\mathbf{u}_\theta
    \right)
    =
    0.
\end{equation}

We augment the flow-matching objective with an $\ell_1$ penalty on the magnitude of the latent transport velocity:
\begin{equation}
    \mathcal{L}(\theta)
    =
    \mathcal{L}_{\mathrm{CFM}}^{\ell_q}(\theta)
    +
    \lambda\,
    \mathbb{E}_{t,\mathbf{z}_t}
    \left[
    \left\|
    \mathbf{u}_\theta(\mathbf{z}_t,t)
    \right\|_1
    \right],
    \qquad
    q\in\{1,2\},
\end{equation}
where $\lambda\geq0$ controls the strength of the penalty. The regularizer directly discourages large latent velocities. It does \emph{not} contain derivatives of $\mathbf{u}_\theta$ and therefore should not be interpreted as a Dirichlet-energy penalty or as direct mathematical control of the Lipschitz constant. Unlike kinetic-energy and Jacobian-norm regularizers for continuous-time flows~\cite{finlay2020train}, which penalize $\|\mathbf u_\theta\|_2^2$ and estimate $\|\nabla_{\mathbf z}\mathbf u_\theta\|_F^2$, respectively, the $\ell_1$ penalty uses only the network output, adds negligible cost, and requires no stochastic trace estimator.

To determine whether reducing velocity magnitude is associated with a smoother learned vector field, we separately evaluate the Jacobian
$\nabla_{\mathbf{z}}\mathbf{u}_\theta$ after training. We report the empirical proxy
\begin{equation}
    \hat{J}
    =
    \max_{(\mathbf{z},t)\in\mathcal{S}}
    \left\|
    \nabla_{\mathbf{z}}
    \mathbf{u}_\theta(\mathbf{z},t)
    \right\|_F ,
\end{equation}
where $\mathcal{S}$ is the finite evaluation set used in the regularization sweep. A decrease in $\hat{J}$ indicates lower local sensitivity over the sampled region, but does not constitute a global Lipschitz bound. Figure~\ref{fig:lambda_sweep} shows that increasing $\lambda$ is empirically associated with a decrease in this Jacobian-norm proxy. The same sweep also shows that this reduction is beneficial only over an intermediate range: excessive regularization suppresses the transport sufficiently strongly that distributional fidelity collapses.

Flow matching is closely related to continuous probability transport, and optimal transport provides useful mathematical context for such dynamics. The Benamou--Brenier formulation of the Wasserstein-2 distance minimizes an $L^2$ kinetic action of the form
\begin{equation}
    \int_0^1
    \int
    \left\|
    \mathbf{u}(\mathbf{z},t)
    \right\|_2^2
    p_t(\mathbf{z})
    \,d\mathbf{z}\,dt
\end{equation}
subject to the continuity equation and prescribed endpoint distributions. The $\ell_1$ velocity penalty used in FlowMesh is a different objective and does not imply recovery of the Benamou--Brenier minimizer or a Wasserstein-2 geodesic. We therefore use optimal transport only as conceptual context for continuous distribution transport, not as a theoretical characterization of the learned FlowMesh trajectories.

\section*{Acknowledgments}
This work used the Delta and DeltaAI systems at the National Center for Supercomputing Applications [awards OAC 2005572 and OAC 2320345] through allocation CIS240093 from the Advanced Cyberinfrastructure Coordination Ecosystem: Services \& Support (ACCESS) program, which is supported by National Science Foundation grants \#2138259, \#2138286, \#2138307, \#2137603, and \#2138296.

\section*{Declaration of Generative AI and AI-Assisted Technologies}
Generative AI tools were used during manuscript preparation for language editing, structural organization, and refinement of selected visual materials. All numerical experiments, reported quantitative results, scientific interpretations, references, and conclusions were independently checked and approved by the authors. The authors take full responsibility for the accuracy, integrity, and originality of the final manuscript.

%Bibliography
\bibliographystyle{unsrt}  
\bibliography{references}  

@inproceedings{finlay2020train,
  title={How to train your neural ode: the world of jacobian and kinetic regularization},
  author={Finlay, Chris and Jacobsen, J{\"o}rn-Henrik and Nurbekyan, Levon and Oberman, Adam},
  booktitle={International conference on machine learning},
  pages={3154--3164},
  year={2020},
  organization={PMLR}
}

@article{dhariwal2021diffusion,
  title={Diffusion models beat gans on image synthesis},
  author={Dhariwal, Prafulla and Nichol, Alexander},
  journal={Advances in neural information processing systems},
  volume={34},
  pages={8780--8794},
  year={2021}
}

@article{tang2026contractive,
  title={Contractive diffusion probabilistic models},
  author={Tang, Wenpin and Zhao, Hanyang},
  journal={SIAM Journal on Imaging Sciences},
  volume={19},
  number={2},
  pages={1364--1389},
  year={2026},
  publisher={SIAM}
}

@inproceedings{yuzhe2025manifold,
  title={Manifold constraint reduces exposure bias in accelerated diffusion sampling},
  author={Yuzhe, YAO and Chen, Jun and Huang, Zeyi and Lin, Haonan and Wang, Mengmeng and Dai, Guang and Wang, Jingdong},
  booktitle={The Thirteenth International Conference on Learning Representations},
  year={2025}
}

@article{kobayashi2024deep,
  title={Deep neural operator-driven real-time inference to enable digital twin solutions for nuclear energy systems},
  author={Kobayashi, Kazuma and Alam, Syed Bahauddin},
  journal={Scientific reports},
  volume={14},
  number={1},
  pages={2101},
  year={2024},
  publisher={Nature Publishing Group UK London}
}

@article{yao2024uncertainty,
  title={Uncertainty-aware deep learning for reliable health monitoring in safety-critical energy systems},
  author={Yao, Yuantao and Han, Te and Yu, Jie and Xie, Min},
  journal={Energy},
  volume={291},
  pages={130419},
  year={2024},
  publisher={Elsevier}
}

@book{zienkiewicz2005finite,
  title={The finite element method set},
  author={Zienkiewicz, Olgierd C and Taylor, Robert Leroy},
  year={2005},
  publisher={Elsevier}
}

@book{anderson1995computational,
  title={Computational fluid dynamics},
  author={Anderson, John David and Wendt, John and others},
  volume={206},
  year={1995},
  publisher={Springer}
}

@book{quarteroni1994numerical,
  title={Numerical approximation of partial differential equations},
  author={Quarteroni, Alfio and Valli, Alberto},
  year={1994},
  publisher={Springer}
}

@article{benner2015survey,
  title={A survey of projection-based model reduction methods for parametric dynamical systems},
  author={Benner, Peter and Gugercin, Serkan and Willcox, Karen},
  journal={SIAM review},
  volume={57},
  number={4},
  pages={483--531},
  year={2015},
  publisher={SIAM}
}

@book{frey2007mesh,
  title={Mesh generation: application to finite elements},
  author={Frey, Pascal Jean and George, Paul-Louis},
  year={2007},
  publisher={Iste}
}

@article{thompson1984grid,
  title={Grid generation techniques in computational fluid dynamics},
  author={Thompson, Joe F},
  journal={AIAA journal},
  volume={22},
  number={11},
  pages={1505--1523},
  year={1984}
}

@article{mavriplis1999directional,
  title={Directional agglomeration multigrid techniques for high-Reynolds-number viscous flows},
  author={Mavriplis, Dimitri J},
  journal={AIAA journal},
  volume={37},
  number={10},
  pages={1222--1230},
  year={1999}
}

@book{thompson1998handbook,
  title={Handbook of grid generation},
  author={Thompson, Joe F and Soni, Bharat K and Weatherill, Nigel P},
  year={1998},
  publisher={CRC press}
}

@ARTICLE{brostein2017geometric,
  author={Bronstein, Michael M. and Bruna, Joan and LeCun, Yann and Szlam, Arthur and Vandergheynst, Pierre},
  journal={IEEE Signal Processing Magazine}, 
  title={Geometric Deep Learning: Going beyond Euclidean data}, 
  year={2017},
  volume={34},
  number={4},
  pages={18-42},
  doi={10.1109/MSP.2017.2693418}}

@InProceedings{sanchez2020learning,
  title = 	 {Learning to Simulate Complex Physics with Graph Networks},
  author =       {Sanchez-Gonzalez, Alvaro and Godwin, Jonathan and Pfaff, Tobias and Ying, Rex and Leskovec, Jure and Battaglia, Peter},
  booktitle = 	 {Proceedings of the 37th International Conference on Machine Learning},
  pages = 	 {8459--8468},
  year = 	 {2020},
  editor = 	 {III, Hal Daumé and Singh, Aarti},
  volume = 	 {119},
  series = 	 {Proceedings of Machine Learning Research},
  month = 	 {13--18 Jul},
  publisher =    {PMLR},
  url = 	 {https://proceedings.mlr.press/v119/sanchez-gonzalez20a.html}
}

@inproceedings{rombach2022high,
  title={High-resolution image synthesis with latent diffusion models},
  author={Rombach, Robin and Blattmann, Andreas and Lorenz, Dominik and Esser, Patrick and Ommer, Bj{\"o}rn},
  booktitle={Proceedings of the IEEE/CVF conference on computer vision and pattern recognition},
  pages={10684--10695},
  year={2022}
}

@article{ho2020denoising,
  title={Denoising diffusion probabilistic models},
  author={Ho, Jonathan and Jain, Ajay and Abbeel, Pieter},
  journal={Advances in neural information processing systems},
  volume={33},
  pages={6840--6851},
  year={2020}
}

@article{lipman2022flow,
  title={Flow matching for generative modeling},
  author={Lipman, Yaron and Chen, Ricky TQ and Ben-Hamu, Heli and Nickel, Maximilian and Le, Matt},
  journal={arXiv preprint arXiv:2210.02747},
  year={2022}
}

@article{lipman2024flow,
  title={Flow matching guide and code},
  author={Lipman, Yaron and Havasi, Marton and Holderrieth, Peter and Shaul, Neta and Le, Matt and Karrer, Brian and Chen, Ricky TQ and Lopez-Paz, David and Ben-Hamu, Heli and Gat, Itai},
  journal={arXiv preprint arXiv:2412.06264},
  year={2024}
}

@inproceedings{qi2017pointnet,
  title={Pointnet: Deep learning on point sets for 3d classification and segmentation},
  author={Qi, Charles R and Su, Hao and Mo, Kaichun and Guibas, Leonidas J},
  booktitle={Proceedings of the IEEE conference on computer vision and pattern recognition},
  pages={652--660},
  year={2017}
}

@article{lu2021learning,
  title={Learning nonlinear operators via DeepONet based on the universal approximation theorem of operators},
  author={Lu, Lu and Jin, Pengzhan and Pang, Guofei and Zhang, Zhongqiang and Karniadakis, George Em},
  journal={Nature machine intelligence},
  volume={3},
  number={3},
  pages={218--229},
  year={2021},
  publisher={Nature Publishing Group UK London}
}

@article{he2024sequential,
  title={Sequential deep operator networks (s-deeponet) for predicting full-field solutions under time-dependent loads},
  author={He, Junyan and Kushwaha, Shashank and Park, Jaewan and Koric, Seid and Abueidda, Diab and Jasiuk, Iwona},
  journal={Engineering Applications of Artificial Intelligence},
  volume={127},
  pages={107258},
  year={2024},
  publisher={Elsevier}
}

@article{jin2022mionet,
  title={MIONet: Learning multiple-input operators via tensor product},
  author={Jin, Pengzhan and Meng, Shuai and Lu, Lu},
  journal={SIAM Journal on Scientific Computing},
  volume={44},
  number={6},
  pages={A3490--A3514},
  year={2022},
  publisher={SIAM}
}

@misc{goswami2022physicsinformeddeepneuraloperator,
      title={Physics-Informed Deep Neural Operator Networks}, 
      author={Somdatta Goswami and Aniruddha Bora and Yue Yu and George Em Karniadakis},
      year={2022},
      eprint={2207.05748},
      archivePrefix={arXiv},
      primaryClass={cs.LG},
      url={https://arxiv.org/abs/2207.05748}, 
}

@article{li2020fourier,
  title={Fourier neural operator for parametric partial differential equations},
  author={Li, Zongyi and Kovachki, Nikola and Azizzadenesheli, Kamyar and Liu, Burigede and Bhattacharya, Kaushik and Stuart, Andrew and Anandkumar, Anima},
  journal={arXiv preprint arXiv:2010.08895},
  year={2020}
}

@article{li2023geometry,
  title={Geometry-informed neural operator for large-scale 3d pdes},
  author={Li, Zongyi and Kovachki, Nikola and Choy, Chris and Li, Boyi and Kossaifi, Jean and Otta, Shourya and Nabian, Mohammad Amin and Stadler, Maximilian and Hundt, Christian and Azizzadenesheli, Kamyar and others},
  journal={Advances in Neural Information Processing Systems},
  volume={36},
  pages={35836--35854},
  year={2023}
}

@inproceedings{hao2023gnot,
  title={Gnot: A general neural operator transformer for operator learning},
  author={Hao, Zhongkai and Wang, Zhengyi and Su, Hang and Ying, Chengyang and Dong, Yinpeng and Liu, Songming and Cheng, Ze and Song, Jian and Zhu, Jun},
  booktitle={International conference on machine learning},
  pages={12556--12569},
  year={2023},
  organization={PMLR}
}
\section{Appendix}

\subsection{Random Vectors}
Let $(\Omega, \mathcal{F}, \mathbb{P})$ be a probability space.  
A \textit{random vector} in $\mathbb{R}^d$ is a measurable function
\[
X : \Omega \to \mathbb{R}^d
\]
such that for every Borel set $B \in \mathcal{B}(\mathbb{R}^d)$, the event
\[
\{\omega \in \Omega : X(\omega) \in B\} \in \mathcal{F}.
\]
The law (or distribution) of $X$ is the push-forward measure
\[
\mu_X(B) = \mathbb{P}(X \in B) = \mathbb{P}(\{\omega \in \Omega : X(\omega) \in B\}).
\]
If $\mu_X$ is absolutely continuous with respect to the Lebesgue measure $\lambda$ on $\mathbb{R}^d$, then there exists a probability density function (pdf) $p_X$ such that
\[
\mu_X(B) = \int_B p_X(x) \, dx, \quad \forall B \in \mathcal{B}(\mathbb{R}^d).
\]

Expectation is defined for an integrable random vector $X$ as
\[
\mathbb{E}[X] = \int_{\Omega} X(\omega) \, d\mathbb{P}(\omega) = \int_{\mathbb{R}^d} x \, p_X(x)\, dx.
\]

\subsection{Conditional density and Expectation}
Let $(X,Y)$ be a random vector on $\mathbb{R}^m \times \mathbb{R}^n$ with joint density $p_{X,Y}(x,y)$.  
The \textit{marginal densities} are given by
\[
p_X(x) = \int_{\mathbb{R}^n} p_{X,Y}(x,y)\, dy, \quad p_Y(y) = \int_{\mathbb{R}^m} p_{X,Y}(x,y)\, dx.
\]
If $p_Y(y) > 0$, then the \textit{conditional density} of $X$ given $Y=y$ is defined as
\[
p_{X|Y}(x|y) = \frac{p_{X,Y}(x,y)}{p_Y(y)}.
\]

The \textit{conditional expectation} of $X$ given $Y=y$ is
\[
\mathbb{E}[X|Y=y] = \int_{\mathbb{R}^m} x \, p_{X|Y}(x|y)\, dx.
\]
More generally, conditional expectation is defined as a random variable $Z = \mathbb{E}[X|Y]$ satisfying:
\begin{enumerate}
    \item $Z$ is $\sigma(Y)$-measurable, where $\sigma(Y)$ is the $\sigma$-algebra generated by $Y$.
    \item For every bounded measurable $g$, 
    \[
    \mathbb{E}[X g(Y)] = \mathbb{E}[Z g(Y)].
    \]
\end{enumerate}

\subsection{Diffeomorphisms and push-forward maps}
Let $M$ and $N$ be smooth manifolds. A map $T: M \to N$ is a \textit{diffeomorphism} if:
\begin{enumerate}
    \item $T$ is bijective,
    \item $T$ is smooth,
    \item $T^{-1}$ is smooth.
\end{enumerate}

If $\mu$ is a measure on $M$, then the \textit{push-forward measure} under $T$ is
\[
T_\# \mu (B) = \mu(T^{-1}(B)), \quad \forall B \subseteq N.
\]

If $X \sim \mu$ and $Y = T(X)$, then $Y \sim T_\#\mu$.  
If $\mu$ admits density $p_X$ and $T$ is differentiable, then the density of $Y$ is
\[
p_Y(y) = p_X(T^{-1}(y)) \, \Big|\det \nabla T^{-1}(y)\Big|.
\]

\subsection{Velocity fields and Flows}

A vector field on a smooth manifold $M$ is a smooth section of the tangent bundle,
that is, a mapping
\[
V : M \to TM
\]
such that $V(x) \in T_x M$ for all $x \in M$.

Let $u : [0,1] \times \mathbb{R}^d \to \mathbb{R}^d$ be a time-dependent vector field.
The associated flow $\{\psi_t\}_{t \in [0,1]}$ is defined as the solution to the ordinary
differential equation
\[
\frac{d}{dt}\psi_t(x) = u\bigl(t,\psi_t(x)\bigr), 
\qquad \psi_0(x) = x.
\]
If $u$ is measurable in $t$ and Lipschitz continuous in $x$, then the flow exists and is
unique, and each map $\psi_t : \mathbb{R}^d \to \mathbb{R}^d$ is a diffeomorphism.

In flow matching, the vector field is parameterized by a neural network
$u_\theta(t,x)$.
The generative model is defined by the probability flow ordinary differential equation.
\[
\frac{d}{dt} x_t = u_\theta(t,x_t), 
\qquad x_0 \sim p_0,
\]
where $p_0$ is a prescribed reference distribution.
In the idealized setting, the terminal random variable satisfies $x_1 \sim p_1$,
where $p_1$ denotes the target data distribution.

\subsection{Constructing probability paths}
Let $p_0$ and $p_1$ be two probability measures on $\mathbb{R}^d$. A \textit{probability path} is a family $\{p_t\}_{t \in [0,1]}$ interpolating between them, i.e.
\[
p_t \big|_{t=0} = p_0, \quad p_t \big|_{t=1} = p_1.
\]

One natural choice is the linear interpolation in sample space:
\[
x_t = (1-t) x_0 + t x_1, \quad x_0 \sim p_0, \, x_1 \sim p_1.
\]
This induces a probability path $p_t = \text{Law}(x_t)$.  

The instantaneous velocity of this interpolation is
\[
u_t(x_t | x_1) = \frac{x_1 - x_t}{1-t}, \quad t \in [0,1).
\]

Alternatively, one can define optimal transport paths. If $T$ is the optimal transport map sending $p_0$ to $p_1$, then the McCann interpolation is
\[
p_t = \big((1-t)\text{Id} + tT\big)_\# p_0.
\]
This path minimizes the quadratic Wasserstein distance $W_2(p_t,p_s)$ and provides a geodesic in Wasserstein space.  

Flow matching also constructs continuous probability paths, but a generic flow-matching path need not be an optimal-transport geodesic. In the present implementation, training uses straight-line interpolation between independently sampled source and target latent states. The learned velocity field is therefore trained to reproduce that prescribed probability path rather than to solve the Monge--Kantorovich or Benamou--Brenier optimization problem. The connection to optimal transport is conceptual unless the coupling and training objective are explicitly chosen to enforce optimal-transport structure.

\end{document}